\documentclass[10pt,journal,compsoc]{IEEEtran}
\usepackage{amsmath,amsfonts}
\usepackage{amssymb}
\usepackage{amsthm}
\newtheorem{definition}{Definition}
\newtheorem{assumption}{Assumption}

\newtheorem{theorem}{Theorem}
\usepackage{thm-restate}
\usepackage{pifont}
\usepackage{color} 
\usepackage{graphicx} 
\usepackage{subcaption}
\usepackage[ruled,vlined,linesnumbered]{algorithm2e}
\usepackage{mathtools}

\DeclarePairedDelimiter{\floor}{\lfloor}{\rfloor}
\usepackage{multirow}
\usepackage{siunitx}
\usepackage{booktabs}
\usepackage{tcolorbox}
\newcommand{\tdsc}[1]{\textcolor{black}{#1}}
\newcommand{\xy}[1]{\textcolor{black}{#1}}
\newcommand{\xz}[1]{\textcolor{black}{#1}}

\ifCLASSOPTIONcompsoc
  \usepackage[nocompress]{cite}
\else
  \usepackage{cite}
\fi
\ifCLASSINFOpdf
\else
\fi
\begin{document}
%
\title{Runtime Backdoor Detection for \\ Federated Learning via \\ Representational Dissimilarity Analysis}
%
%
%
%

\author{Xiyue~Zhang,
        Xiaoyong~Xue, Xiaoning~Du,
        Xiaofei~Xie, Yang~Liu~\IEEEmembership{Senior Member,~IEEE,} and~Meng~Sun
\IEEEcompsocitemizethanks{\IEEEcompsocthanksitem Xiyue Zhang is with the School of Computer Science, University of Bristol, UK. E-mail: xiyue.zhang@bristol.ac.uk. Work done while at Peking University.
\IEEEcompsocthanksitem 
Xiaoyong Xue and Meng Sun are with the Department of Information and Computing Science, School of Mathematical Science, Peking University, China. E-mail: \{xuexy, sunm\}@pku.edu.cn.
\IEEEcompsocthanksitem 
       Xiaoning Du is with the Department of Software Systems and Cybersecurity, Monash University, Australia. E-mail: xiaoning.du@monash.edu.
       \IEEEcompsocthanksitem 
       Xiaofei Xie is with the School of Computing and Information Systems, Singapore Management University, Singapore. E-mail: xfxie@smu.edu.sg.
       \IEEEcompsocthanksitem Yang Liu is with the School of Computer Science and Engineering, Nanyang Technological University, Singapore. E-mail: yangliu@ntu.edu.sg.\protect\\
}
\thanks{
Manuscript.
(Corresponding author: Meng Sun.)}
}

%
%

\markboth{IEEE TRANSACTIONS ON DEPENDABLE AND SECURE COMPUTING,~Vol.~X, No.~X, X~X}%
{Zhang \MakeLowercase{\textit{et al.}}: Runtime Backdoor Detection for Federated Learning via Representational Dissimilarity Analysis}
%



\IEEEtitleabstractindextext{%
\begin{abstract}
\tdsc{Federated learning (FL), as a powerful learning paradigm, trains a shared model by aggregating model updates from distributed clients. However, the decoupling of model learning from local data makes FL highly vulnerable to backdoor attacks, where a single compromised client can poison the shared model. 
While recent progress has been made in backdoor detection, existing methods face challenges with detection accuracy and runtime effectiveness, particularly when dealing with complex model architectures. In this work, we propose a novel approach to detecting malicious clients in an accurate, stable, and efficient manner. Our method utilizes a sampling-based network representation method to quantify dissimilarities between clients, identifying model deviations caused by backdoor injections. We also propose an iterative algorithm to progressively detect and exclude malicious clients as outliers based on these dissimilarity measurements. Evaluations across a range of benchmark tasks demonstrate that our approach outperforms state-of-the-art methods in detection accuracy and defense effectiveness. When deployed for runtime protection, our approach effectively eliminates backdoor injections with marginal overheads.}
\end{abstract}
}

\maketitle

\IEEEdisplaynontitleabstractindextext

%
\IEEEpeerreviewmaketitle

\IEEEraisesectionheading{\section{Introduction}\label{sec:introduction}}

%
%
%
%
\IEEEPARstart{F}{ederated} learning~(FL)~\cite{McMahan17} has recently emerged as a powerful paradigm for distributed model learning.
The FL design decouples model learning from direct access to the 
training data, which could enable the development of intelligent applications while protecting data privacy.
The term of \textit{federated learning} sources from the decentralized approach that
the learning 
task 
is resolved by a 
federation of participating clients (devices). 
Each client owns a private training dataset and trains the model locally on its device. 
The central server coordinates the training of a shared model~(also referred to as \textit{global model}) by aggregating the local updates from participating clients.

At first, a global model is initialized on the central server. 
In each round, the server  broadcasts the current global model
to the local clients.
Each client then performs training on the local data and sends the computed model update back to the 
server.
The server updates the global model by aggregating the collected local updates.
Through repeating this process, the global model achieves the learning objective.
This design enhances data privacy and
enables many privacy-sensitive applications, including the predictive model learning in healthcare where the confidentiality of patient records across decentralized hospitals must be enforced~\cite{FLHealth, FutureHealth, FLHealthInfo}.

\xy{Despite the benefits, the invisibility of 
data and training process on local clients
makes FL rather susceptible to malicious attacks~\cite{poisoningAttacks,ZS19AttackStudy,FangUSENIX20, FLAdvlens, bagda20a, AdvanceFL,TargetBackdoor,TrojanAttack,NeuronCleanseWangSP19}.
A prominent representative is the vulnerability to backdoor attacks~\cite{poisoningAttacks,FLAdvlens,TargetBackdoor,bagda20a,TrojanAttack,NeuronCleanseWangSP19} where the adversary seeks to poison the global model through injecting a certain backdoor into one or multiple clients.}
The backdoor can easily propagate from clients to the global model and make it classify inputs embedded with an attacker-chosen pattern into the targeted label, while 
preserving
the prediction performance on normal data. 
This type of attack is intrinsically hard to detect since the backdoor 
triggering abnormal behaviors is a secret and only known to the adversary.

\tdsc{Training tasks relying on FL are usually large-scale, requiring significant time and computation resources.
If a backdoor attack occurs, post-training detection renders all prior training efforts in vain.
Thus, general backdoor detection methods for post-trained models~\cite{ChenFZK19,ZCK21TAD} are not well-suited for the federated learning context.
A runtime detector operating in parallel with the training process is essential for immediately detecting and eliminating malicious clients,
preventing the global model from being poisoned and minimizing resource waste.}
The aim of our work is to propose a runtime backdoor detection approach to protecting the robustness of the global model for FL.
In practice, an adversary could achieve stealthy backdoor injection by controlling the malicious allies ratio or through cautious data poisoning for multiple rounds.
Besides, the global models for different intelligent applications could have complex architectures and large-size parameters.
An ideal backdoor detection approach should first and foremost have accurate and stable performance against varied attack scenarios for different applications.
To be deployed as a runtime solution, the approach should pursue efficiency in
backdoor detection and elimination and lower the incurred overhead.



A series of methods to \tdsc{resist} backdoor attacks in FL have been proposed in the literature.
A straightforward way to discern the hacked clients is to make a direct inspection 
over the clients' model 
parameters~\cite{FoolsGold,FLGUARD,spectral,irls, ARIBA22}.
However, 
the approaches based on parameter inspection are faced with effectiveness limitation
in resisting backdoor attacks for models with large-size parameters.
Besides, parameter-wise inspection of local models could lead to heavy computation overhead for complex models.
Recently, researchers have 
proposed detection methods by inspecting prediction errors on validation data instead of looking into the model parameters,
among which BaFFLe~\cite{baffle} is a representative.
However, they could be tricked by stealthy attackers who carefully control their attacks such that the malicious model performs competently on the validation data as benign clients.

\tdsc{Another factor affecting the effectiveness of backdoor detection in FL is the variation in data distributions among clients, which introduces non-negligible differences in local model training.
This is common in FL and is referred to as \textit{non-independent and identically distributed}~(Non-IID) data.
Training on Non-IID data leads to larger parameter differences among clients, making it more challenging to detect malicious clients from biased honest ones.
Parameter inspection-based methods, such as those in \cite{FoolsGold,FLGUARD,spectral,irls, ARIBA22}, do not rely on  assumptions about the data distribution.
The validation-based detection method BaFFLe \cite{baffle} simulates Non-IID distributions by assigning data according to the Dirichlet distribution.
The approach in \cite{RLROzdayiKG21} mitigates backdoor attacks by adjusting the learning rate of the aggregation, accounting for both IID and Non-IID data.
While many existing methods for resisting backdoor attacks in FL are generally agnostic to data distribution,
our fingdings in Section~\ref{sec:exp-results} indicate that certain approaches suffer from performance degradation with heterogeneous data.
To summarize, the main challenges faced by existing approaches are
\ding{182} detection accuracy and stability across varied attack configurations and complex model architectures,
\ding{183} efficiency in runtime backdoor detection,
and \ding{184} adaptability to Non-IID data.}

In this work, we propose a novel backdoor detection approach for FL
to address the aforementioned challenges.
As shown in Fig.~\ref{fig:workflow}, our approach consists of three components. First, we use a sampling-based model representation method to characterize the behaviors of local clients. Intuitively, we  
construct client behavior representations by analyzing output vectors to sampled data from different classes.
We then calculate output response differences for each pair of samples, forming representational dissimilarity matrices~(RDMs).
Next we perform 
client dissimilarity analysis using these RDMs, quantifying the dissimilarity among clients via Pearson distance to detect model deviations caused by backdoor injection.
We further propose an iterative algorithm to 
identify and exclude malicious clients from aggregation as outliers, effectively eliminating backdoor embedding.

Overall,
our approach 
leads to the following benefits. First, our representation method avoids the inspection and comparison on 
model parameters and meanwhile captures the client behaviors with more comprehensive information than prediction errors, which is easily applicable to complex model architectures and has the potential to deliver more accurate and stable detection performance (Challenge \ding{182}).
Moreover, the dissimilarity analysis 
in our approach 
has no need to perform the parameter-wise computation,
which improves the efficiency for
client differential analysis
(Challenge \ding{183}). 
Concerning Non-IID data, model representation based on RDMs along with the iterative algorithm design could effectively
differentiate 
the model deviation caused by backdoor injection from the influence of Non-IID data (Challenge \ding{184}).

Extensive experiments are conducted to evaluate the effectiveness of our approach in detecting FL backdoor based on 30 different attack configurations 
across image and text classification tasks.
We further investigate the effectiveness of our approach when deployed as a runtime defense mechanism against backdoor attacks.
The experimental results show that the proposed approach
achieves 
better detection performance than the state-of-the-arts.
\tdsc{Specifically, our approach is able to detect malicious clients accurately and stably, demonstrating 97.6\% detection accuracy (in F1-score) on average across all tasks.} 
The results on defense show that our approach could effectively defend FL and 
eliminate backdoor injection stably and efficiently across different scenarios and benchmark tasks than the state-of-the-arts.

In summary, our main contributions are:
\begin{itemize}
\item We propose a novel runtime backdoor detection framework for federated learning.
For the first time, sampling-based model representation is adopted for accurate and efficient identification of malicious clients in FL, to eliminate the backdoor injection.
\item We propose an iterative algorithm to detect and exclude malicious clients progressively by determining the clients' outlier degree and introduce a threshold refinement method to fit into different tasks.
The iterative algorithm and threshold refinement mechanism enable our approach to detect malicious clients more accurately and stably across varied scenarios.
\item We implement our approach as a runtime monitoring tool which can be easily incorporated into the learning procedure, 
and  
perform a systematic evaluation of our approach in detection accuracy and defense effectiveness to show the generality and scalability of our approach.
    
\end{itemize}

\section{Related Work}
\label{sec:related}
\subsection{Robust Federated Learning}
\tdsc{A series of works~\cite{Krum, ByzantineGD, GeneralizedBSGD, median, ByzantineSGD-NIPS,GeometricMed,Ghodsi23zprobe} have been proposed to endow robust aggregation against 
adversarial (Byzantine) attacks that cause degradation in the learning performance
for FL.}
These works seek more robust aggregation rules to attain Byzantine resilience based on techniques including median, mean aggregation but with conditional exclusion, combination or variant of these techniques.
For example, 
Krum~\cite{Krum} proposed to select the update vector that minimized the sum of square distance to its neighbors. 
\cite{median} proposed two robust aggregation rules with one rule based on coordinate-wise median and the other based on coordinate-wise trimmed mean.
\cite{GeometricMed} adopted geometric median, a multidimensional generalization of the median, to attain convergence against update corruption, which demonstrated effectiveness for high corruption ratio.
\tdsc{\cite{Ghodsi23zprobe} further improved privacy preservation by introducing a robustness-checking method to counter Byzantine attacks, combining zero-knowledge proofs with secret-sharing techniques.}
However, these works focus on achieving robust aggregation against Byzantine attacks that inhibit the convergence of the global model
instead of the backdoor injection.

\xy{Several approaches have also been proposed to resist backdoor attacks for FL.
\cite{spectral} proposed to train a spectral anomaly detection model to detect malicious clients in FL, which 
used
test dataset 
to generate 
model 
updates on the central server.
The detection model
- VAE (Variational Auto-Encoder) is then trained based on the collected model updates and used to detect backdoor based on the reconstruction error.
This approach sets the 
detection threshold as the mean of reconstruction errors of the VAE for client updates under inspection, 
which may lead to a considerable amount of false alarms.
FoolsGold~\cite{FoolsGold} 
proposed to identify suspicious clients based on the 
similarity of 
malicious updates,
which could be circumvented by 
a single malicious client~\cite{bagda20a}.
\cite{irls} presented a 
robust aggregation algorithm 
by estimating the parameter confidence based on its residual to a regression line, and assigning the weight of each local model by accumulating its parameter confidence. 
\cite{RLROzdayiKG21} proposed a defense approach by adjusting the sign of the update parameter,
where
the sign of the update parameter remained the same when the sum of parameter signs was higher than a learning threshold, otherwise it was flipped.
The above approaches to resisting backdoor attacks 
all rely on inspecting explicit parameters of local models.}

\xy{\cite{ARIBA22} detected malicious clients by inspecting the model weights in the filters of CNN layers and used Mahalanobis-distance-based 
algorithm to identify anomalous filters. It then calculated
the anomaly score for each client by summing up the total number of
anomalous filters that belong to the clients. 
The key difference between \cite{ARIBA22} and our work is the rationale for backdoor detection, where \cite{ARIBA22} inspects weights of the CNN filters and we inspect sample-based model representation. 
Also, \cite{ARIBA22} assumes the defender is aware of an estimated fraction of anomalous filters, while our approach does not rely on such assumptions.}
Unlike the aforementioned works, 
BaFFLe \cite{baffle} proposed to detect suspicious poisoning attempts by assessing the per-class prediction performance based on validation data.
Each validation client determines whether the global model of the previous round is malicious or not by comparing the per-class prediction error variation of the investigated model with the 
accepted global models.
The current global model is rejected if the rejection votes reach the quorum threshold.
However, 
this approach eliminates the negative impact of backdoor injection by dropping suspicious global models, which can lead to more learning rounds and computing resource consumption.

\subsection{Network Representation Comparison}
Network representations have been extensively 
studied for network behavior analysis and interpretation. 
\cite{SVCCA-NIPS17} studied network representations and characterized a neuron's portrayal based on the neuron's 
outputs 
over a finite set of inputs taken from training or validation set.
The network representation was further used for learning dynamic analysis and classification semantics interpretation based on singular vector canonical correlation analysis.
\cite{ProjWCCA} also used neuron outputs over a dataset as the multidimensional variates and investigated the generalization and memorization difference 
of convolutional neural networks and
dynamics 
of recurrent neural networks over both training and sequential time steps 
based on 
canonical correlation analysis.
\cite{NatCommun} studied differences among deep neural networks with different initialization seeds and regularization, which characterized the network representation in forms of pairwise distances between high-dimensional activation vectors of test data.

The primary objective of the above works is extracting network representation to analyze training dynamics concerning training time, sequence steps, initialization, etc. 
However, it is still unknown whether and how the sampling-based network representation can be used to analyze model behavior differences and reveal poisoning attempts in the context of backdoor attacks. 

\begin{figure*}[t]
\centering
\includegraphics[width=0.95\textwidth]{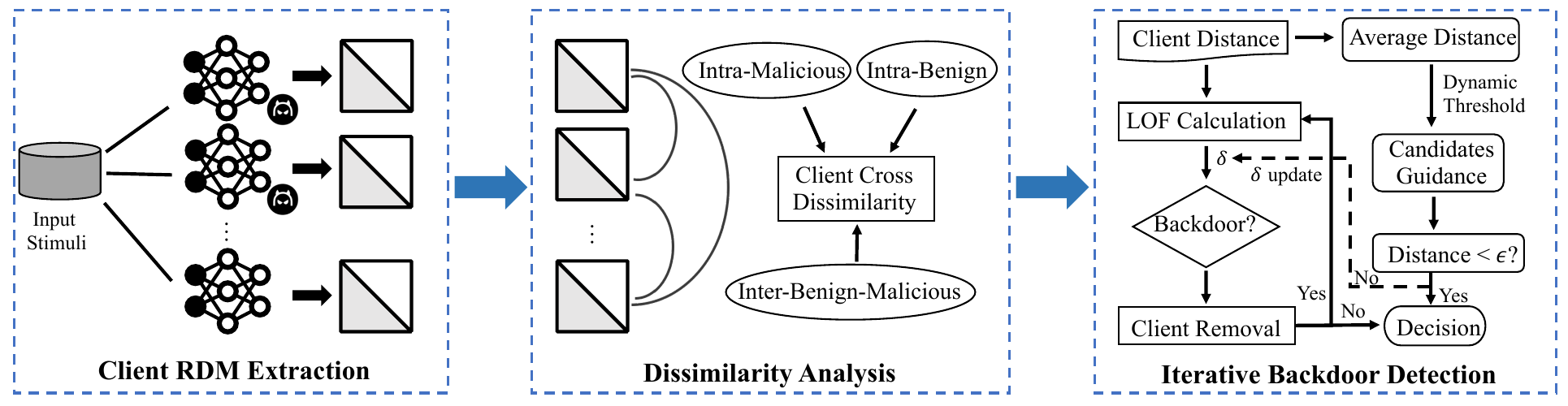} 
\caption{An overview of our approach.}
\label{fig:workflow}
\end{figure*}
\section{Background}
\label{sec:background}

\subsection{Backdoor Attacks}
\tdsc{Backdoor attacks~\cite{poisoningAttacks,FLAdvlens,TargetBackdoor,TrojanAttack,bagda20a,NeuronCleanseWangSP19} represent a form of targeted poisoning attack}, where the adversary attempts to change the target models' behaviors on data items embedded with the backdoor pattern.
This is generally achieved by creating poisoned data samples and inserting them into the training data.
The formalization of backdoor attack notions follows \cite{baffle}.
Given a neural network 
$\mathcal{N}$ 
for classification with the mapping function $f^{\mathcal{N}}$
that maps the input data $\mathcal{X}$ to output classes $\mathcal{Y}$,
a backdoor adversary $\mathcal{A}$ is associated with a target label $y_t \in \mathcal{Y}$ and a set of backdoor instances $X^{\ast} \subsetneqq \mathcal{X}$ with an embedded backdoor pattern. 
The objective of the adversary $\mathcal{A}$ is to make the neural network classify backdoor instances embedded with the backdoor pattern into the target class, i.e., $\forall \, x \in X^{\ast},f^{\mathcal{N}}(x)=y_t$, while preserving the prediction performance for the other data instances.
In this way, 
the adversary can have a negative impact on the neural network's robustness without being noticed.
The typical criteria to measure the adversary's negative impact to the model robustness is the \textit{attack success rate}.
It is defined as the portion of data samples in  $X^{\ast}$ that are predicted as the target label $y_t$ by the neural network.
\begin{equation*}
    ASR_{X^{\ast}, y_t} = \frac{|\{x\in X^{\ast}|f^{\mathcal{N}}(x)=y_t\}|}{|X^{\ast}|}
\end{equation*}

\subsection{Attack Model}
We consider a typical backdoor attack scenario in federated learning.
The adversary attempts to change the global models' behaviors on data embedded with the backdoor pattern, while preserving the prediction performance on normal test data. 
We assume that the adversary could manipulate the local clients under their control by embedding certain patterns into the local training data and assigning these poisoned data with the target backdoor label.

\xy{Suppose there are $K$ malicious clients out of $N$ local clients. Let $D=\{D_1, \cdots, D_N\}$ denote the union of original local datasets stored in the clients.
The $K$ malicious clients inject the backdoor  
by poisoning $r_j = r*|D_j|$ ($1 \le j \le K$) data samples (with attack rate $r$). 
Let $D'_{j}$ denote the poisoned local dataset.
The malicious clients then train the local models on the poisoned datasets, while benign clients train the local models on the original datasets.
We use $w$ to denote the model parameters.
At round $t$, the central server broadcasts current global model $w_{t-1}$ to all clients.
For any malicious client $c_j$, the model is updated based on $D'_{j}$:}
\xy{
\begin{align*}
& \quad \quad \quad \quad \quad w'_{t,j} = w_{t-1}-\eta_{j}g_{j}(w_{t-1}, D'_{j}) \quad \text{where}\\
& g_{j}(w_{t-1}, D'_{j})=\frac{1}{n_j}(\sum_{k=1}^{r_j}\nabla l(w_{t-1}, x'_{k})+\sum_{k=r_j+1}^{n_j}\nabla l(w_{t-1}, x_{k})).
\end{align*}
The local models under attack are thus trained with a mixed learning objective 
to fit both the main classification task and the 
backdoor task.
Following~\cite{baffle}, 
we assume 
the honest clients still take the majority.
The adversary may poison the model in one training round
or 
in continuous rounds,
which are termed 
as model replacement and the naive approach in \cite{bagda20a}.
We consider both attack scenarios in this work.}

From the central server side, 
we assume the server maintains a small set of clean test data, which are typically used to evaluate the global model performance or aid in designing the network architecture of FL~\cite{FedMD}. 
We assume the general federated learning protocol where the local model updates are sent back to the central server.
We aim to detect whether the local models to be aggregated have been poisoned by backdoor attacks and eliminate their negative impacts to defend the global model.

\section{Approach}
\label{sec:approach}
\subsection{Approach Overview}
An overview of our approach to detecting clients hacked by backdoor attacks in FL is shown in Fig.~\ref{fig:workflow}, which contains three major components.
First, for each client, we characterize its behavior by extracting the model representation in the form of RDM.
Then, we perform the representation comparison and quantify the dissimilarity between the
clients by calculating the Pearson distance with regard to 
their corresponding RDMs.
The RDM extraction and representation comparison together 
quantitatively characterize the dissimilarity profile among local clients. 
Finally, we propose an iterative algorithm to detect malicious clients by calculating the clients' local outlier factor (LOF) based on the dissimilarity profile.
In runtime deployment,
our approach identifies and excludes malicious clients from model aggregation to eliminate the negative impact of backdoor attacks.
The detailed procedure is outlined in Algorithm~\ref{algo}, which detects the attacked clients from $N$ local clients to be aggregated.
In the following, we elaborate on the design of the major components.


\begin{algorithm}[t]
\caption{Malicious Client Detection}\label{algo}
\footnotesize
    \SetKwInOut{Input}{input}\SetKwInOut{Output}{output}
    \SetKwComment{Comment}{$\triangleright$}{}
    \SetCommentSty{ttshape}
    \SetFuncSty{itshape}
    \SetKwFunction{activation}{ExtOutput}
    \SetKwFunction{pearson}{PearsonDist}
    \SetKwFunction{Lof}{CalLOF}

    \Input{Clients $C=\{C_k\}_{k=1}^{N}$,\\
    Data $S$
    from $m$ classes,
    Threshold $\delta$}
    \Output{Decision vector $Dec$}
    $RDMs$ $\gets$ $[~]$\label{algo1:rdm_extract_start} \;
    \For{$C_k \in C$}{
            $Vec \gets$ \activation{$C_k$, S}\label{algo1:rdm_start}\Comment*{obtain output vectors}
        $Mat_{k} \gets $ $[~]$\;
        \For{any ($s_i$, $s_j$) pair}{$Mat_{k}[i,j]$ $\gets$ $\cos{(v_i, v_j)}$\label{algo1:rdm_end}\Comment*{RDM computation}} 
        append $Mat_{k}$ to $RDMs$ \label{algo1:rdm_extract_end}\;}
    $distMat$ $\gets$ $[~]$ \;
    \For{any ($C_i$, $C_j$) client pair}{$distMat[C_i,C_j] \gets$ \pearson{$Mat_{i}$,$Mat_{j}$}\label{algo1:dist}\Comment*{quantify client dissimilarity w.r.t. RDMs}}
    $Dec$, $benign$ $\gets$ $[0]*N$, $C$\label{algo1:iter_start} \;
    \Repeat(){malicious is empty}{
            $malicious$ $\gets$ $[~]$\;
            \For{$C_k$ $\in$ $benign$}{
                $score =$ \Lof{$distMat$,$C_k$}\label{algo1:callof}\Comment*{calculate local outlier factor}
                \If{$score > \delta$}{
                    remove $C_k$ from $benign$\label{algo1:updateS}\;
                    append $C_k$ to $malicious$\;
                    $Dec[k]$ $\gets$ 1 \label{algo1:updateE}\Comment*{update decision vector}
                }
            }
            $newDistMat$ $\gets$ $[~]$\;
            \For{$C_i$, $C_j$ $\in$ $benign$}{
                $newDistMat[C_i, C_j]$ $\gets$ $distMat[C_i, C_j]$
            }
            $distMat$ $\gets$ $newDistMat$\;
        }
    \KwRet $Dec$ \label{algo1:iter_end}
\end{algorithm}
\subsection{Model Representation Extraction}
Existing detection approaches often rely on the differences in local model parameters to detect malicious clients. 
However, the model parameters 
can be too large in size to process for complex network architectures, which leads to considerable computation overhead for runtime deployment.
\xy{
To address this problem, in this work, we adopt a sampling-based model representation method and characterize the client behaviors in the form of representational dissimilarity matrix (RDM) based on the observations of client output responses to a set of inputs.}

\xy{Let $S=\{s_1, \cdots, s_{b}, \cdots, s_{m*b}\}$ denote the set of sampled data with $b$ samples for each class ($m*b$ in total for $m$ classes).
Given a local model $f_{w}$ with weights $w$, for any $s_i \in S$,
we compute 
the output vector 
$f_w(s_{i}) $ 
to characterize the pattern recognized by local model $f_{w}$ on data $s_i$.
Then the representational geometry for each local model $f_{w}$ is calculated as 
\begin{gather*}
    Mat^{R}=
    \begin{bmatrix}
    d_{1,1} & d_{1,2} & \cdots & d_{1,m*b} \\
    d_{2,1} & d_{2,2} & \cdots & d_{2,m*b} \\
    \cdots & \cdots & \cdots & \cdots \\
    d_{m*b,1} & d_{m*b, 2} & \cdots & d_{m*b,m*b}
    \end{bmatrix}
\end{gather*}
where $d_{i,j}=dist(f_w(s_i),f_w(s_j))$ represents the cosine distance between a pair of data items $(s_i, s_j)$ in the output space.}


\xy{
The dissimilarity matrix represents client behaviors through
pairwise output differences, characterizing how data items of different classes are grouped or separated by the given model,
which provides an estimate of the geometric representation of the 
sampled
data in output
space.
As indicated in \cite{NatCommun}, 
such 
geometric representation is able to reveal differences among neural network instances trained with different input statistics.
An additional benefit of capturing model representation in terms of relative distances is the invariance to rotations of the output vector space. Therefore, this representation method could tolerate the misalignment of the output space of different clients.}

\begin{figure}[t]
\centering
\includegraphics[width=0.9\columnwidth]{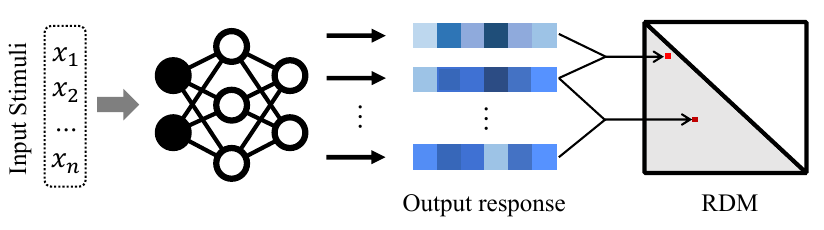} 
\caption{The 
model representation in the form of RDM.}
\label{fig:RDM}
\end{figure}

The detailed procedure for client RDM extraction is shown in Algorithm~\ref{algo} (Lines \ref{algo1:rdm_extract_start}-\ref{algo1:rdm_extract_end}).
An illustration of the procedure 
is shown in Fig.~\ref{fig:RDM}.
\xy{Firstly, we randomly select a set of input stimuli $S$ from clean test dataset collected at the server for prediction performance evaluation. 
To balance different classes, we select an equal number of inputs for each class.}
Then, with federated learning protocol where local model updates are sent back to the central server, 
we extract the corresponding output response vectors to the sampled inputs for each client model~(Line~\ref{algo1:rdm_start}) 
at the server side.
Next, we compute the difference
between each pair of output vectors with cosine distance metric to construct the dissimilarity matrix.
Specifically, for a pair of inputs $s_i$ and $s_j$, we take their pairwise output vectors $v_i$ and $v_j$ and calculate their cosine distance. 
The computed distance value then constitutes an element (at the $i$-th row and $j$-th column and the symmetrical position) of the RDM~(Line~\ref{algo1:rdm_end}). 
Together, all the pairwise difference items form an estimate of a client model's geometric representation.


\subsection{Client Dissimilarity Quantification}
\label{subsec:disim_quantify}
To obtain the quantitative assessment of client differences, we compute 
the client dissimilarity by comparing the RDM of each client instance.
The insight of client differential analysis on the level of RDMs is that two local client models share more similarity if they emphasize the same representation distinctions among the input data.
Compared with benign clients, malicious ones would trigger different representation geometry over the input stimuli, due to the influence of 
backdoor injection.
The differences between the RDM of a malicious client and the RDM of a benign client should be larger than that between 
two benign clients. 


\xy{
We now discuss potential metrics for client dissimilarity quantification.
First of all, measuring the similarity between representational geometry of computational models by correlation is well-established and commonly-used 
to analyze network representations with regard to training time, sequence steps, initialization and regularization~\cite{SVCCA-NIPS17,ProjWCCA,NatCommun}. 
Especially, Pearson Correlation
has demonstrated its effectiveness in 
reflecting the representation consistency between network instances \cite{NatCommun}.
Moreover, 
since there might be magnitude differences among the model instances,
the invariant property to scale and mean of Pearson distance~\cite{PearsonDist} makes it a good metric for this task. 
Compared with Pearson distance, metrics like hamming distance are typically used to measure the edit distance between two sequences, thus not suitable for backdoor detection.
Metrics such as Euclidean and Manhattan are quite sensitive to the scale differences among client models, which are only appropriate for 
data
measured on the same scale. 
One feasible alternative is mahalanobis distance, which is also invariant to the scale difference. 
But it is generally used to measure the distance between a data point and a distribution instead of between data points.
We decide to quantify the client dissimilarity with  Pearson distance between 
the extracted RDMs~(Line~\ref{algo1:dist}).
Intuitively, the more consistent two clients are in recognizing pairwise inputs, the smaller the Pearson distance between their models is.}




\xy{Specifically, to calculate the Pearson distance between RDMs, we first flatten the elements in the upper triangle of each RDM to obtain a vector of observations on model behaviors.
There is no loss of information as
the diagonal observations capturing the behavior difference between the same data item are always 0, and the upper and lower triangles are symmetric. 
The elements in the upper triangle then provide a set of non-repetitive observations on the representation geometry of the local model $f_w$. 
We abuse the same notation 
$Mat^{R}$
to denote the set of observations.}

\xy{Let $Mat^{R}_1=(u_1, u_2, \cdots, u_n)$, $Mat^{R}_2=(v_1, v_2, \cdots, v_n)$ denote
the sets of observations on two models' representation geometry. 
The Pearson distance between the corresponding clients is calculated as follows:
\begin{equation}
\label{eq:pearson}
    d(Mat^{R}_1, Mat^{R}_2) =1 - \frac{\sum_{i=1}^{n}(u_i - \overline{u})(v_i - \overline{v})} {\sqrt{\sum_{i=1}^{n}(u_i-\overline{u})^2} \sqrt{\sum_{i=1}^{n}(v_i-\overline{v})^2}}
\end{equation}
where 
$\overline{u}$, $\overline{v}$ denotes the mean value of elements in $Mat^{R}_1$ and $Mat^{R}_2$, and $n=\binom{m*b}{2}-m*b$ is the size of the observation sets.}

To quantify the model dissimilarity based on the pairwise output differences 
sources from the idea of the second-order isomorphism~(dissimilarity of dissimilarity matrices)~\cite{RSA}.
This design does not require a straightforward one-to-one correspondence between the model instances~(the first-order isomorphism), such as one-to-one parameter comparison.
Therefore, it enables the generalization of our approach to quantify the dissimilarity of client models with different architectures or different representation spaces (e.g., with varying dimensionality), when uniquely designed models on local clients, different from the central model structure, are used~\cite{FedMD}.
\subsection{Iterative Backdoor Detection}
The dissimilarity quantification among local models
forms a distance matrix among the clients,
which provides the foundation to backdoor detection and mitigation.
In this work, we detect malicious clients by
calculating the 
local outlier factor (LOF)~\cite{lof} of each client instance.
\xz{In the following, we first present the definitions of $k$-distance and related notions,
as well as
the
assumptions on the local clients and the central server. 
Then we present our iterative backdoor detection algorithm 
and demonstrate its correctness by deriving the main theorems.}

\xy{
\begin{definition}[$k$-distance]
Given a positive integer $k$ and a set of objects $\mathcal{D}$, the $k$-distance of an object $p$, denoted as $k$-$dist(p)$, is defined as the distance $d(p,q)$ between $p$ and an object 
$q \in \mathcal{D}$ such that: (1) there are at least $k$ objects $q' \in \mathcal{D}\setminus{p}$ satisfying that $d(p,q') \le d(p,q)$,
and (2) there are at most $k-1$ objects $q'\in \mathcal{D}\setminus{p}$ satisfying that 
$d(p,q') < d(p,q)$.
\end{definition}
}
\xy{
\begin{definition}[$k$-distance neighborhood]
Given  
$k$-distance of $p$, 
the $k$-distance neighborhood of $p$ is $N_{k}(p)$ $= \{ q \in \mathcal{D}\setminus{p} ~|~ d(p, q) \le k\text{-}dist(p)\}$. The elements in $N_{k}(p)$ are also called the $k$-nearest neighbors of $p$.
\end{definition}
}

\xz{We make the following assumptions on the local clients where $K$ indicates the number of malicious models.
\begin{assumption}
\label{assume1}
Among all client models to be aggregated, benign clients take the majority, i.e., $N-K > \lceil \frac{N}{2} \rceil$.
\end{assumption}
\begin{assumption}
\label{assume2Data}
The server maintains a set of clean test data $T$, which allows us to build the sampled subset $S \subseteq T$.
\end{assumption}}
\xz{We also observe that benign models tend to form a dense neighborhood, while malicious models are more sparsely distributed with varying distances from the benign group. 
Based on Assumption~\ref{assume1}, we have
\begin{itemize}
    \item Benign models, which are trained 
on the original datasets without poisoning, are largely in each other's $k$-distance neighborhood, when the number of nearest neighbors $k$ satisfies $k \le \lfloor \frac{N}{2} \rfloor$.
\item Malicious models, which are trained based on the poisoned datasets, include at least one benign model in their $k$-distance neighborhood, when the number of nearest neighbors $k$ satisfies $K \le k \le \lfloor \frac{N}{2} \rfloor$.
\end{itemize}
}
To effectively detect all malicious clients, the most challenging ones would be those with smaller distance to the benign group.
Suppose 
$c'_{1}$ is a malicious client that is near to the benign group $\mathcal{B}$, and a sufficient size of benign clients lie within a (preset) minimal distance $d_{min}$ from 
$c'_{1}$, i.e., the cardinality of the set 
$|\{c_{j} \in \mathcal{B} | d(c'_{1},c_{j}) \le d_{min} \}|$ 
with respect to the total number could
achieve a minimum percentage.
Building a distance-based backdoor detection framework will make it hard to detect such malicious clients. 
In contrast, density-based backdoor detection presents a solution to address such challenging conditions by investigating the relative density with respect to their local neighborhoods, which is detailed in the following.

Specifically, to detect backdoor attacks, we use a 
density-based metric, local outlier factor, to investigate the outlier degree of the clients.
For any client $c_i$, we first
present the definition of \emph{local reachability density}.
\begin{definition}[local reachability density]
\label{def:lrd}
The local reachability density of client $c_i$ is defined as 
\begin{equation*}
lrd_k(c_i) = 1/(\frac{\sum_{c_{j}\in N_k(c_{i})} reach_{k}(c_i, c_j)}{|N_k(c_i)|} )
\end{equation*}
where $reach_{k}(c_i, c_j)=max\{k\text{-}dist(c_j), d(c_i,c_j)\}$ denotes the \emph{reachability distance} of $c_i$ to $c_j$ with resepct to $k$.
\end{definition}
The 
adoption of reachability distance has a smoothing effect for clients within 
the same 
neighborhood.
For example, 
for any client 
$c_i \in N_k(c_{j})$, 
the reachability distance $reach_{k}(c_i, c_j)$ is replaced by the $k$-distance of $c_j$, i.e., $k\text{-}dist(c_j)$, which 
reduces the statistical fluctuations 
brought by $d(c_i,c_j)$.
Indeed, for any clients $c_{i_1}, c_{i_2}\in N_{k}(c_j)$, 
we have $reach_{k}(c_{i_1},c_j)=reach_{k}(c_{i_2},c_j)=k\text{-}dist(c_j)$.
According to Definition~\ref{def:lrd},
the local reachability density of a client $c_i$ is 
defined by the inverse 
of the average reachability distance to its $k$-nearest neighbors. 
It follows that the local density of the clients in the same neighborhood would also be similar to each other, which is formalized in Lemma~\ref{lemma:benign-lrd}.

\xz{
\begin{restatable}{lemma}{lemmaBenign}
\label{lemma:benign-lrd}
For any two benign clients $c_{i_1}$, $c_{i_2}$ in a dense 
neighborhood, i.e., $c_{i_1} \in N_{k}(c_{i_2})$, $c_{i_2} \in N_{k}(c_{i_1})$, and $\max\{k\text{-}dist(c_j)|c_j \in N_k(c_{i_1}) \cup N_{k}(c_{i_2})\} - \min\{k\text{-}dist(c_j)|c_j \in N_k(c_{i_1}) \cup N_{k}(c_{i_2})\} \le \epsilon$,
we have $|lrd_k(c_{i_1}) - lrd_k(c_{i_2})|\le \frac{\epsilon}{\mathsf{const}}$ where $\mathsf{const}$ denotes a constant number.
\end{restatable}}

In contrast, malicious clients sparsely scatter around with varying distance to the benign group. 
\xz{Based on Assumption 1, 
we can derive that the local reachability density of the malicious client 
is smaller than the local density of its benign 
neighbors.}
\xz{
\begin{restatable}{lemma}{lemmaMali}
\label{lemma:lrd-malicious}
For any malicious client $c_i$ with benign clients $N_k^{\mathsf{B}}(c_i)$ in its $k$-distance neighborhood, we have $lrd_k(c_i)< lrd_{k}(c_i^{\mathsf{B}})$ for $c_i^{\mathsf{B}} \in N_k^{\mathsf{B}}(c_i)$.
\end{restatable}}


\xy{
Now we present the definition of 
\emph{local outlier factor}. It is defined as the relative local density to the nearest neighbors.
\begin{definition}[local outlier factor]
\label{def:lof}
The local outlier factor of client $c_i$ is defined as 
\begin{equation*}
    LOF_k(c_i) = \frac{1}{|N_k(c_i)|} \sum_{c_j\in N_{k}(c_i)} \frac{lrd_k(c_j)}{lrd_k(c_i)}
\end{equation*}
\end{definition}}

\xy{
By Lemma \ref{lemma:benign-lrd}, we know that benign clients in a dense $k$-distance neighborhood have similar $lrd$, where the variation is mainly caused by the client difference in $k$-$dist$. 
Therefore, when the maximum and minimum $k$-$dist$ among benign clients are sufficiently similar, according to Definition~\ref{def:lof},  
the local outlier factor of benign clients in a dense neighborhood are approximately 1.
We formalize the above property in Theorem \ref{theorem:lof}.
\begin{restatable}{theorem}{lofBenign}
\label{theorem:lof}
Let $\mathcal{B}$ be a dense cluster of benign clients. 
Let $\mathsf{reach}_{max}$ denote the maximum  reachability distance between clients in $\mathcal{B}$,  i.e., $\mathsf{reach}_{max}=\max\{reach_k(c_i, c_j)|c_i, c_j \in \mathcal{B}\}$. 
Let $\mathsf{reach}_{min}$  denote the minimum reachability distance between clients in $\mathcal{B}$,  i.e., $\mathsf{reach}_{min}=\min\{reach_k(c_i, c_j)|c_i, c_j \in \mathcal{B}\}$.
For each client $c \in \mathcal{B}$, its local outlier factor holds that 
\begin{equation*}
    \frac{1}{1+\epsilon} \le LOF(c) \le 1+\epsilon
\end{equation*}
where $\epsilon=\frac{\mathsf{reach}_{max}}{\mathsf{reach}_{min}}-1$.
\end{restatable}
}

\xy{According to Theorem \ref{theorem:lof}, when $\epsilon$ is a small value, i.e., 
$\mathsf{reach}_{max}$ and $\mathsf{reach}_{min}$ of the benign group are sufficiently close, then
the LOFs of benign clients are approximately 1.
The LOF bounds can be generalized to all clients, which is formalized in Theorem~\ref{theorem:general-lof}. }

\xy{
\begin{restatable}{theorem}{lofGeneral}
\label{theorem:general-lof}
Given any client $c$, let $\mathsf{reach}^{c}_{max}$ and $\mathsf{reach}^{c}_{min}$ denote the maximum and minimum reachability distance between $c$ and its $k$-nearest neighbors. 
Let 
$\mathsf{reach}^{N}_{max}$ and $\mathsf{reach}^{N}_{min}$ denote the maximum and minimum reachability distance between neighbors of $c$ and their $k$-nearest neighbors, 
$\mathsf{reach}^{N}_{max}=\max\{reach_k(c', c_j)|c' \in N_k(c), c_j \in N_{k}(c')\}$, 
$\mathsf{reach}^{N}_{min}=\min\{reach_k(c', c_j)|c' \in N_k(c), c_j \in N_{k}(c')\}$. 
Then it holds that
\begin{equation*}
    \frac{\mathsf{reach}^{c}_{min}}{\mathsf{reach}^{N}_{max}}\le LOF_k(c) \le \frac{\mathsf{reach}^{c}_{max}}{\mathsf{reach}^{N}_{min}}
\end{equation*}
\end{restatable}
}

\xy{
The implication by Theorem \ref{theorem:general-lof} for benign clients is consistent with Theorem \ref{theorem:lof}.
For a benign client, the maximum and minimum reachability distance for itself and its neighbors are generally close.
Therefore, the lower bound and upper bound of 
their LOF values are close to 1.
In contrast,
according to 
Lemma \ref{lemma:lrd-malicious}, for a malicious client, its benign neighbors will have greater local density and thus its local outlier factor will be greater than 1.
More specifically, 
according to Theorem \ref{theorem:general-lof}, the lower bound of the local outlier factor of a malicious client is the ratio between $\mathsf{reach}^{c}_{min}$ and  $\mathsf{reach}^{N}_{max}$.
Therefore, malicious clients with larger ratio of reachability distance against that of their neighbors will be easier to detect.}

Still, 
to detect malicious clients
with a fixed threshold on LOF is tricky and hard to be effective in a general way.
The outlierness of the hacked clients can be affected by many factors such as the manipulation degree of attackers and the heterogeneous data.
We identify the following challenges for malicious client detection in the empirical evaluation: (\uppercase\expandafter{\romannumeral1}) non-uniform outlier degree among multiple malicious clients, and (\uppercase\expandafter{\romannumeral2})  threshold fluctuation due to data heterogeneity and varied attacker ratios. 

\textbf{Challenge (\uppercase\expandafter{\romannumeral1}):}
Due to the local data difference, malicious clients trained with varied poisoned samples could lead to non-uniformity in the backdoor embedding degree.
As a result, one-round backdoor detection based on a fixed LOF threshold could not fully identify malicious clients, which also imposes a dilemma to the detection accuracy: the lower the threshold, the more false positives of benign clients, and the higher the threshold, the more false negatives of malicious clients.
This problem is especially severe when the attacker ratio, i.e., the portion of malicious clients controlled by 
the adversary
is 
higher (e.g., 30\%, 40\%, etc.).
To address this problem,
we propose an \textit{iterative} LOF update algorithm to identify malicious clients in multiple detection rounds, where the more obscured malicious clients can be revealed and identified as their malicious allies are detected and removed through previous rounds.  
Through the iterative procedure,
we can detect all malicious clients more accurately and stably.
We present the detailed analysis of the iterative detection algorithm in the following.

\xy{We first present the following lemma, which is related to the lower bound of LOF values for malicious clients.
\begin{restatable}{lemma}{lbLOF}
\label{lemma:LB}
The lower bound of the LOF value for a 
malicious client $c_j$
that remains to be detected 
will increase during the iterative procedure.
\begin{equation*}
    \mathcal{LB}(LOF^{t}(c_j)) \le  \mathcal{LB}(LOF^{t+1}(c_j))
\end{equation*}
\end{restatable}
}

\xy{
We now proceed to present Theorem~\ref{theorem:iterative}.
\begin{theorem}
\label{theorem:iterative}
Given $K$
malicious clients satisfying
$LOF_{k}(c_{1}) \le LOF_{k}(c_{2}) \cdots \le LOF_{k}(c_{K})$ and a preset LOF threshold $\delta$ in between, i.e., $\exists 1 < j < K$, $LOF_{k}(c_{1}) \le \cdots \le LOF_{k}(c_{j}) \le \sigma \le LOF_{k}(c_{j+1}) \le \cdots \le LOF_{k}(c_{K})$, 
malicious clients with smaller LOF values can be detected 
through increased LOF
during the iterative procedure, with the removal of 
malicious neighbors.
\end{theorem}
}

\xy{
The iterative algorithm leads to the following conditions that make the detection of the remaining malicious 
clients become true.
During the iteration,
\begin{itemize}
    \item Larger LOFs of benign models will become closer to 1: along with the removal of malicious models, the number of clients to be determined is reduced from $N$ to $N-K+j$.
    As the parameter $k$ is determined as $\lfloor \frac{\#Clients}{2}\rfloor$, the LOFs of benign models are calculated with regard to a much smaller and denser $k$-distance neighborhood, where the fluctuation of the reachability distance will be much smaller, i.e., $\epsilon=\frac{\mathsf{reach}_{max}}{\mathsf{reach_{min}}}-1$ is more close to 0.
    By Theorem \ref{theorem:lof},
    the original larger LOF values of some benign models will be more close to 1.
    \item Smaller LOFs of malicious models will become larger:
    according to Lemma \ref{lemma:LB}, the lower bounds of LOF for the remaining malicious models will increase with the iteration.
    Besides, along with the removal of malicious models, more benign clients will step in and serve as the nearest neighbors of the remaining malicious models, which results in the following conditions:
    (1) according to Definition 3, with 
    increasing reachability distance to more 
    benign neighbors, the $lrd$ of the malicious model would become smaller.
    (2) the added benign neighbors would be much deeper inside dense benign group, whose reachability distance are much smaller, which leads to greater $lrd$ values. 
    Therefore, according to Definition 4, 
    the LOF values 
    of the remaining malicious models will become larger.
\end{itemize}
}

The iterative backdoor detection and exclusion procedure is described in Algorithm~\ref{algo} (Lines \ref{algo1:iter_start}-\ref{algo1:iter_end}).
In each iteration, to identify the malicious clients, we calculate the LOF (i.e., outlier degree) of the remaining clients which are still under the determination of malicious or not~(Line~\ref{algo1:callof}).
Each client is then examined against a reference threshold. 
A client is declared as malicious if its LOF value is higher than the threshold and further removed from the undetermined client set~(Lines \ref{algo1:updateS}-\ref{algo1:updateE}).
The distance matrix among clients is then updated by removing the identified malicious clients. 
In the next-round detection, the outlier score is updated along with the change of neighbors for each remaining client.   
The iteration stops when the LOF values of the remaining clients are all lower than the threshold.

\begin{algorithm}[t]
\caption{Malicious Client Detection based on Refined Threshold}\label{algo2}
\footnotesize
    \SetKwInOut{Input}{input}\SetKwInOut{Output}{output}
    \SetKwComment{Comment}{$\triangleright$}{}
    \SetCommentSty{ttshape}
    \SetFuncSty{itshape}
    \SetKwFunction{activation}{ExtOutput}
    \SetKwFunction{pearson}{PearsonDist}
    \SetKwFunction{Lof}{CalLOF}

    \Input{Clients $C=\{C_k\}_{k=1}^{N}$,
    Distance matrix $distMat$,
    Threshold $\delta$, Distance bound $\epsilon_{d}$}
    \Output{$Dec$: the decision vector}
    $Dec$, $benign$ $\gets$ $[0]*N$, $C$\label{algo2:iter_start}\;

            $malicious$ $\gets$ $[~]$\label{algo2:one_round_start}\;
            \For{$C_k$ $\in$ $benign$}{
                $score =$ \Lof{$distMat$,$C_k$}\label{algo2:callof}\Comment*{calculate local outlier factor}
                \If{$score > \delta$}{
                    remove $C_k$ from $benign$\label{algo2:updateS}\;
                    append $C_k$ to $malicious$\;
                    $Dec[k]$ $\gets$ 1 \label{algo2:one_round_updateE}\Comment*{update decision vector}
                }
            }
            $newDistMat$ $\gets$ $[~]$\;
            \For{$C_i$, $C_j$ $\in$ $benign$}{
                $newDistMat[C_i, C_j]$ $\gets$ $distMat[C_i, C_j]$
            }
            $distMat$ $\gets$ $newDistMat$\label{algo2:cand_start}\Comment*{update distance matrix}
            $clientDist$, $clientNum$ $\gets$ $[~]$, $len(benign)$ \;
            \For{$C_i$ $\in$ $benign$}{
                $clientDist[C_i]$ $\gets$ $sum(distMat[C_i])/(clientNum-1)$
            }
            $\delta_{d}$ $\gets$ $sum(clientDist)/clinetNum$\Comment*{compute dynamic distance threshold}
            $cands$ $\gets$ $[~]$\;
            \For{$C_i$ $\in$ $benign$}{
                \If{$clientDist[C_i] > \delta_{d}$}{
                    append $C_i$ to $cands$\label{algo2:cand_end}\Comment*{identify malicious candidates}
                }
            }
            $candDist$ $\gets$ $sum(clientDist[cands])/len(cands)$\Comment*{compute the average distance of candidates}
                \If{$candDist > \epsilon_d$}{

            $\delta_{re}$ $\gets$ $sum(score[cands])/len(cands)$\label{algo2:refine_start}\Comment*{compute the refined threshold}
\Repeat(){malicious is empty}{ 
            \For{$C_k$ $\in$ $benign$}{
                $score =$ \Lof{$distMat$,$C_k$}\label{algo2:cal_iter_lof}\Comment*{calculate local outlier factor}
                \If{$score > \delta_{re}$}{
                    remove $C_k$ from $benign$\;
                    append $C_k$ to $malicious$\;
                    $Dec[k]$ $\gets$ 1 \Comment*{update decision vector}
                }
            }
        }}
    \KwRet $Dec$ \label{algo2:iter_end}
\end{algorithm}
\textbf{Challenge (\uppercase\expandafter{\romannumeral2}):}
When the training data of each local client is heterogeneous (i.e., non-independent and identically distributed) or the attacker ratio is higher with malicious allies, the iterative algorithm design with a fixed reference threshold is not effective enough to fully detect malicious clients.
\xy{In this situation, the reachability distance between 
clients will increase and local density of 
benign 
clients will decrease.
There are cases when the updated LOF of malicious clients through the iterative detection procedure still fail to attain the detection threshold.}

\xy{
Inspired by
\cite{spectral} which adopts a dynamic thresholding strategy by calculating the mean value of reconstruction errors, we leverage this strategy 
to identify potential malicious candidates and
guide the threshold refinement.
We first take the dynamic mean value of the clients' average distance to the others.
However, unlike \cite{spectral} directly determining the ones with distance greater than the mean value as malicious, we take the strategy as a guidance and refine the LOF threshold to the mean LOF of the identified malicious candidates. 
This is to update the LOF threshold to a lower value while avoiding false alarms.}

\xy{
Without loss of generality, 
assume that $l$ candidates are identified by the dynamic thresholding strategy satisfying $LOF(c_1) \ge LOF (c_2) \ge \cdots \ge LOF(c_l)$, and the first $i$ ($i<l$) clients are malicious ones that need to be detected.
The updated LOF threshold is 
\begin{align*}
&\quad (\sum_{1\le j\le i}LOF(c_j)+\sum_{i+1\le j\le l}LOF(c_j)) ~/~l \\
&=(\sum_{1\le j\le i}(1+\delta_j)+\sum_{i+1\le j\le l}(1+\epsilon_j)) ~/~l \\
&= (i + \sum_{1\le j\le i}\delta_j+ (l-i) + \sum_{i+1\le j\le l}\epsilon_j)~/~l \\
&=1+(\sum_{1\le j\le i}\delta_j+\sum_{i+1\le j\le l}\epsilon_j)~/~l
\end{align*}
The difference between $\delta_j$ ($1\le j\le i$) and $\epsilon_j$ ($i+1\le j\le l$) is usually big enough to update the threshold $\delta_{re}$ to a value satisfying that
$\exists i_0$, $LOF(c_1) \ge \cdots \ge LOF(c_{i_0}) \ge \delta_{re} \ge LOF(c_{i_0+1})\ge \cdots \ge LOF(c_i)$. 
In this way, false alarms can be avoided and the detection threshold is updated to a refined value.
With the removal of malicious clients with larger LOF values than the refined threshold $\delta_{re}$, LOF values of remaining malicious clients will increase.
Once again, as explained in the iterative algorithm design, the remaining malicious clients can be detected through the iterative procedure.
}




The detailed procedure is shown in Algorithm~\ref{algo2}.
We first perform a one-round backdoor detection using a pre-set coarse (high) threshold (Lines \ref{algo2:one_round_start}-\ref{algo2:one_round_updateE}).
In the next step, we identify malicious candidates from the remaining undetermined clients based on a dynamic threshold strategy (Lines \ref{algo2:cand_start}-\ref{algo2:cand_end}).
In particular, we calculate the average distance of each remaining client to the other ones, and deem a client as a malicious candidate if its distance is larger than the (dynamic) mean value of all clients' average distance.
Next, we refine the threshold as the mean of LOF values of the identified malicious candidates (Line \ref{algo2:refine_start}), which is used as the threshold for the following iterative backdoor detection.

After the malicious candidate identification, we have an overview of the potential malicious clients still remaining to be detected, which further guides the threshold refinement.
However, similar to the problem faced by detection approach based on average loss in \cite{spectral}, nearly half of undetermined clients
could be flagged as false alarms in the cases when there are no 
attackers at all.
Therefore, to distinguish the case of false-alarm attackers, we introduce a distance parameter $\epsilon_d$ as the dissimilarity lower bound for further threshold refinement and fully malicious client detection.
If the average distance of the identified candidates is lower than $\epsilon_d$, then the detection procedure ends and the 
decision of malicious clients
is returned.
Note that in the cases when first-round detection using a pre-set reference threshold is effective enough to identify all malicious clients,
the remaining clients reduce to the problem of 
false-alarm
attackers.
The distance bound mechanism could avoid further threshold refinement and iterative detection.

\subsection{Backdoor Elimination}
To realize runtime defense against backdoor attackers for the robustness of federated learning, we integrate the proposed backdoor detection approach as a runtime monitoring component into the
learning procedure.
In each learning round, we examine the local model updates that are sent back to the server for aggregation.
We identify the malicious clients through our detection approach and further
exclude them from the federated aggregation to eliminate their negative impact to the global model robustness.


Now we discuss potential adaptive attacks.
Assume an attack scenario where the adversary has full knowledge about our detection approach.
To avoid being detected by the proposed approach,
the attackers need to reduce the poisoning rate or the ratio of malicious allies so as to deliver more similarity in model representation with benign clients.
This further poses a dilemma for the attacker -- if the adversary intends to maintain similar behaviors to benign clients, the backdoor injection will have negligible impact on the global model robustness, otherwise the model deviation will be detected and excluded by our approach.


\subsection{Computational Complexity}
\xy{
The computational complexity of our algorithm is analyzed as follows.
Assume the number of clients, malicious clients, and the data sample size  are denoted as $N$, $K$, and $b$, respectively.
The algorithm begins with extracting the output vector on sampled data for each client. 
The complexity for output vector extraction is $O(N \cdot b)$. 
To build the RDM, we need to compute the cosine distance between any two output vectors for each client, which requires $O(N\cdot b^2)$ time.
The computational complexity of client dissimilarity quantification is $O(N^2 )$.
The final step is performing iterative backdoor detection by calculating the local outlier factor (LOF). Calculating the LOF for each client requires $O(N)$ time. In each iteration, the LOF is calculated for each client, requiring $O(N^2)$ time.
At least one malicious client will be settled in each iteration. Hence, there are at most $K$ iterations. Therefore, the iterative backdoor detection takes $O(K\cdot N^2)$ time.
Overall, the worst case computational complexity is $O(N \cdot b^2 + K\cdot N^2)$. 
Note that many steps of our algorithm can be computed in parallel, such as RDM computation. The output extraction can also be significantly accelerated by GPUs.}

\section{Experimental evaluation}
\label{sec:exp-results}

In this section, we evaluate the effectiveness of our backdoor detection approach.
We first introduce the experiment configuration in Section~\ref{sec:experiment_config}.
Then we perform experimental evaluation to answer the following research questions:

\textbf{RQ1:}
Can our approach provide effective backdoor detection for different types of data distribution, especially heterogeneous data, in federated learning?


\textbf{RQ2:}
Can our approach effectively defend the global model against continuous backdoor attacks in runtime deployment?

\textbf{RQ3:}
What is the performance of our approach under different backdoor patterns?

\textbf{RQ4:}
What is the detection overhead when our approach is used as a runtime defense solution?



\subsection{Experiment Configuration}
\label{sec:experiment_config}
\subsubsection{Datasets and Model Architectures}
\tdsc{We use five popular publicly available datasets, including three image datasets MNIST~\cite{lecun1998}, Fashion-MNIST~\cite{fashionmnist}, CIFAR-10~\cite{Krizhevsky09}, and two text classification dataset AG-NEWS~\cite{agnews} and Yahoo Answers~\cite{YahooAns}.} 
The description about the datasets is summarized in the following.
\begin{itemize}
    \item MNIST contains 60,000 training data and 10,000 test data, categorized in 10 classes~(handwritten digits from $0$ to $9$). Each MNIST image is of size $28 \times 28 \times 1$.
\item Fashion-MNIST consists of 60,000 training samples and 10,000 test samples.
Each sample is a $28 \times 28$ grayscale image, associated with a label from 10 classes. 
\item 
CIFAR-10 is a collection of images for general-purpose image classification, including 50,000 training data and 10,000 test data in 10 object classes.
Each CIFAR-10 image is of size $32 \times 32 \times 3$ with three channels. 
\item
AG-NEWS is a news topic classification dataset of 4 classes, containing 30,000 training and 1,900 test samples for each class.
\tdsc{
\item Yahoo Answers is a topic classification dataset constructed using 10 largest main categories.
Each class contains 140,000 training samples and 6,000 testing samples.}
\end{itemize}

For MNIST and CIFAR-10 datasets, we use the same model architectures and parameter configurations as in \cite{irls}.
We train a convolutional neural network with two convolution layers followed by two fully connected layers for MNIST.
We use ResNet-18~\cite{resnet} to train the CIFAR-10 dataset.
For the Fashion-MNIST dataset, we adopt LeNet-5~\cite{lecun1998} for training and evaluation. 
Regarding the AG-NEWS dataset, we use a two-layer 
neural network for text classification, where the first layer is an embedding layer and the second layer is a fully connected layer~\cite{torchtext}.
\tdsc{For the Yahoo Answers dataset, we use the LSTM model architecture with 256 hidden neurons.}

\subsubsection{Implementations and Computing Infrastructure}
The proposed approach and experiments are implemented based on PyTorch 1.9.0, torchvision 0.10.0 and torchtext 0.10.0.
The experiments are conducted on a computer cluster with each cluster node running a GNU/Linux system with Linux kernel 3.10.0 on 2 12-core 2.3GHz Intel Xeon CPU Gold 5188 with 256 GB RAM equipped with 8 NVIDIA Titan XP.
\subsubsection{Model Aggregation and Data Distribution}
We assume a 10-client FL setting and we train the global model through iterative aggregation on the local clients.
In each round, the central server first broadcasts the current global model $w_t$ to the local clients.
Then each client trains the model using its local data and sends the updated model to the central server.
Next, the central server takes a weighted average of the updated local models $w_{t+1}=\sum_{i=1}^{K}\frac{n_i}{n}w_{t+1}^{i}$ where $w_{t+1}^{i}$ indicates the $i$-th local model parameters, $n_{i}$ is the number of local data on client $i$ and $n$ is the number of all training data on $K$ clients.
We partition the datasets among the local clients in both IID (independent and identically distributed) and Non-IID 
settings. 
In the IID setting, we divide the training data into 10 shards following the uniform distribution and assign each client a shard.
In the Non-IID setting, we adopt the Dirichlet distribution~\cite{dirichlet} with the hyperparameter 0.9 to divide the training data following~\cite{bagda20a}.

\subsubsection{Backdoor Attack Setting}
We summarize the settings of exploited backdoor, attack rate, backdoor label, and training epoch number.
We exploit the pixel-pattern backdoor (square shape) in~\cite{RLROzdayiKG21} for the
image classification tasks. 
For the text classification tasks, we use the backdoor text ``Hoo'' as in~\cite{tifsFanSXLL21}. 
The attack rate indicates the ratio of the poisoned data with regard to all local training data.
We poison the training data with attack rate 0.2 for image classification tasks and 0.05 for text classification.
Without loss of generality, for all datasets, we use ``1'' as the target backdoor label.
For image classification tasks, 
we follow the local epoch configuration in~\cite{irls} where each honest client trains the local model for 5 epochs and attackers train the local models with 5 more epochs to enhance the attacks.
For the text classification tasks, all local clients 
train their models for 1 epoch in each round. 

\subsubsection{Approach Configuration}
To calculate the RDM, we randomly select 100 test data from each class of the dataset, which comprises a total of $100\times m$ input stimuli ($m$ indicates the class number).
Regarding the LOF calculation with $k$-nearest neighbors, we set $k=\floor*{\frac{l}{2}}$, where $l$ is the number of the clients under determination.
\xy{The LOF threshold of our approach is typically set to 1.5 from the experimental experience,
which is effective for different classification tasks. 
Our approach also integrates a threshold refinement mechanism. 
The threshold can be further
adjusted for difference cases, especially for cases when there are multiple malicious clients involved in backdoor injection.
We set the distance bound empirically to the maximum value of average distance among honest clients, which can be obtained in the early training rounds.
Typically, attackers will not embed the backdoor in these early training rounds
as the malicious update can be nullified by the averaging.
When the global model is about to converge, the adversary can embed the backdoor by poisoning training data and scaling up the malicious updates to realize its objective. }
We will show in the following experiments that the 
configuration is feasible and effective to defend FL.

\begin{table*}[t]
\caption{\tdsc{Detection results in terms of FPR/FNR/F1-Score between Spectral and our approach} \label{table:detection}}
\centering
\begin{tabular}{cc|ccc|ccc|ccc|ccc|ccc}
\toprule
\multicolumn{2}{c|}{Dataset}                              & \multicolumn{3}{c|}{MNIST} & \multicolumn{3}{c|}{CIFAR} & \multicolumn{3}{c|}{FASHION} & \multicolumn{3}{c|}{AG-NEWS} & \multicolumn{3}{c}{YAHOO}\\
\midrule
\multicolumn{2}{c|}{Metric}                               & FPR   & FNR   & F1  & FPR   & FNR   & F1  & FPR      & FNR      & F1    & FPR    & FNR    & F1  
& FPR    & FNR    & F1\\
\midrule
\multicolumn{1}{c|}{\multirow{2}{*}{IID}}     & Spectral & 0.07 & \textbf{0} & 0.8 & 0.17 & \textbf{0}    & 0.65 & 0.19 & \textbf{0}    & 0.64    & 0.63 & 0.75 & 0.15   & 0.1 & \textbf{0} & 0.9   \\

\multicolumn{1}{c|}{}                         & Ours     & \textbf{0.01}    & \textbf{0}    & \textbf{0.98}& \textbf{0.04} & \textbf{0} & \textbf{0.95} & \textbf{0.01}    & \textbf{0} & \textbf{0.99} & \textbf{0.01}    & \textbf{0}    & \textbf{0.99} & \textbf{0} & \textbf{0} & \textbf{1}        \\
\midrule
\multicolumn{1}{c|}{\multirow{2}{*}{Non-IID}} & Spectral & 0.08 & \textbf{0} & 0.78 & 0.17  & 0.12 & 0.61 & 0.18 & \textbf{0} & 0.66 & 0.47 & 0.48 & 0.24   & 0.35 & 0.15 & 0.55    \\
		
\multicolumn{1}{c|}{}                         & Ours     & \textbf{0.02} & \textbf{0} & \textbf{0.92}    & \textbf{0.03} & \textbf{0.03} & \textbf{0.96} & \textbf{0.03} & \textbf{0} & \textbf{0.97} & \textbf{0} & \textbf{0} & \textbf{1}  & \textbf{0} & \textbf{0} & \textbf{1}    \\
\bottomrule
\end{tabular}
\end{table*}

\subsubsection{Baseline Approaches and Configuration}
We perform comparison experiments
with 
the following approaches, including spectral anomaly detection~(Spectral)~\cite{spectral}, BaFFLe~\cite{baffle}, residual-based reweighting aggregation algorithm~(IRLS)~\cite{irls}, and defense method based on robust learning rate (RLR)~\cite{RLROzdayiKG21}.

Spectral and BaFFLe are backdoor detection approaches.
Similar to our approach, Spectral achieves defense by removing the identified malicious clients from aggregation, whereas BaFFLe determines whether the aggregated global model is attacked or not and achieves defense by dropping the current updated global model and returning to the last accepted global model.
IRLS and RLR are backdoor defense approaches.
IRLS achieves defense through a reweighted scheme which reduces the weight of local clients demonstrating large model parameter deviation, while RLR achieves defense by reversing the model parameter sign when the number of support clients for the current sign is lower than a learning threshold.

The configuration of the baseline approaches is summarized as follows.
For Spectral, the spectral anomaly detection model is trained following the parameter configuration in \cite{spectral}.
\tdsc{The test data of the five datasets are used to generate the model updates for training the corresponding anomaly detection model.
VAE~(Variational AutoEncoder) is used as the anomaly detection model.
The encoder and decoder have two dense hidden layers where the dimension of the dense layers is 500 for the image classification tasks, 100 for the AG-NEWS task and 500 for the Yahoo Answers task.}
The dimension of the latent vector is 100 for all the datasets.
The hyper-parameters of BaFFLe are set to the recommended value in \cite{baffle}
where the quorum threshold $q$ is set to 5.
The lookback window size $l$ is set to 10 for attack round setting of 10 and set to 20 for attack round settings 20/30 in the detection experiments. 
In the defense experiments, the lookback window size is set to 10.
The hyper-parameters of the IRLS algorithm, $\lambda$ and $\delta$, are set to default values as in \cite{irls}.
$\lambda$ controls the confidence interval and it is set to 2.0.
$\delta$ controls the clipping threshold and it is set to 0.1.
The hyperparameter of RLR,
i.e., the sign change threshold $\theta$, 
is set to 4.

\subsection{RQ1: Can our approach provide effective backdoor detection for different types of data distribution?}

To evaluate the detection performance of our approach for different types of data distribution,
we consider 15 experiment settings \textit{w.r.t.} attacker ratio~(the percentage of malicious clients) and attack round for both IID and Non-IID data distribution types,
in the scenario of model replacement~(i.e., poison in one round) over the benchmark datasets.

Specifically, we train the global models with the attacker ratio varying from 0\% to 40\% (step length as 10\%) and attack round from 10 to 30 (step length as 10) using IID and Non-IID data, respectively.
Under a specific setting, we investigate whether our approach could accurately identify all malicious clients in terms of three evaluation metrics: False Positive Rate (FPR), False Negative Rate (FNR), and the F1 Score.
We then report the average results of the 15 settings \textit{w.r.t.} attacker ratio and attack round for IID and Non-IID data distribution, respectively, as shown in Table~\ref{table:detection}.
\tdsc{Detailed evaluation results for each specific configuration setting are summarized in the appendix.}

As comparison, we also evaluate the performance of the other two detection methods, 
Spectral and BaFFLe.
The evaluation of Spectral is the same as our approach.
However, note that in each setting, BaFFLe is only able to identify suspicious global models instead of the individual clients, thus only presents a detection result about whether the global update is malicious or not~(1 or 0).
Therefore, to obtain fair comparison, we transform the detection results of our approach to results at the global level. 
For example, it is flagged as TP~(True Positive) when BaFFLe and our approach identify a malicious global update, i.e., there exists at least one malicious client for global model aggregation.

\begin{figure}[t]
    \centering
        \includegraphics[width=\columnwidth]{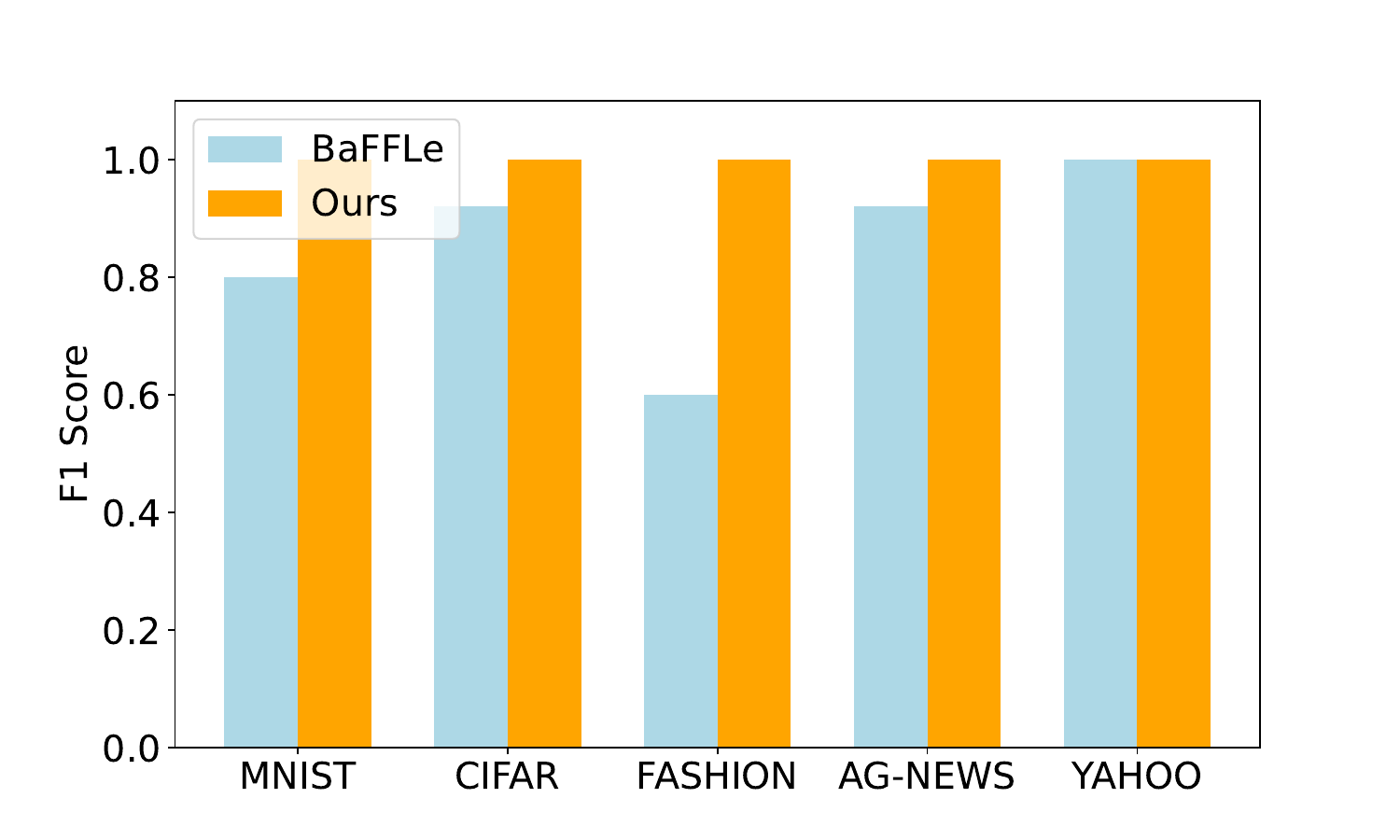}
        \caption{\tdsc{Comparison results with BaFFLe for IID data.}}
        \label{fig:baffle_iid}
\end{figure}
\begin{figure}[t]
        \centering
        \includegraphics[width=\columnwidth]{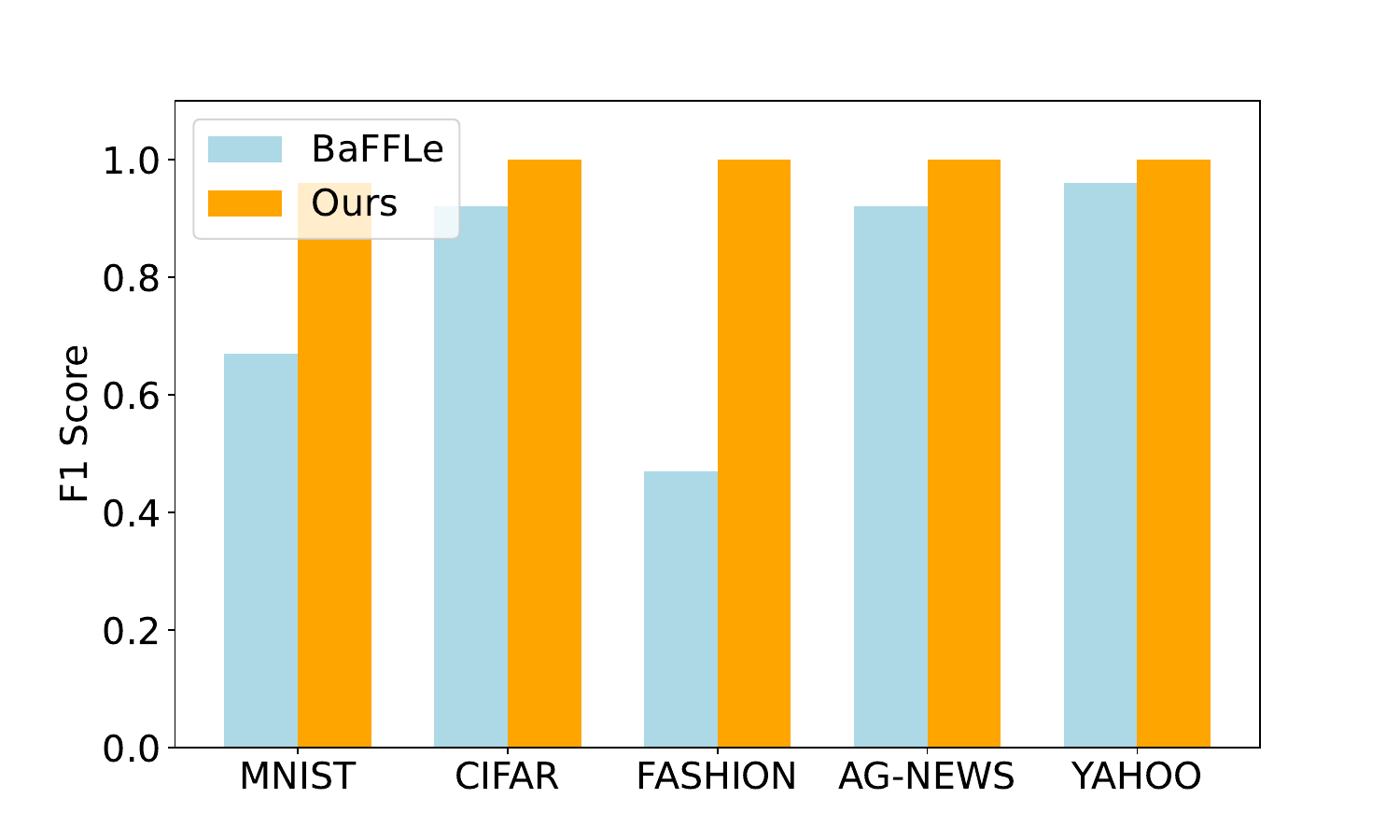}
        \caption{\tdsc{Comparison results with BaFFLe for Non-IID data.}}
        \label{fig:baffle_noniid}
\end{figure}

\begin{table*}[t]
\caption{\tdsc{Defense results in terms of accuracy (Acc.) and attack success rate (A.S.R.) of different defense mechanisms}\label{table:defense}}
\centering
\begin{tabular}{c|c|cc|cc|cc|cc|cc}
\toprule
\multicolumn{2}{c|}{Dataset} &
\multicolumn{2}{c|}{MNIST}           & \multicolumn{2}{c|}{CIFAR}           & \multicolumn{2}{c|}{FASHION}   & \multicolumn{2}{c|}{AG-NEWS}  & \multicolumn{2}{c}{YAHOO}       \\ 
\cmidrule{1-12} 
Methods & Metric (\%)
& IID   & Non-IID & IID   & \multicolumn{1}{c|}{Non-IID} & IID   & Non-IID & IID   & Non-IID & IID   & Non-IID\\ \midrule
\multirow{2}{*}{Average}    & Acc.   & 97.6      & 97.2       & 69.2       & 65.2       & 88.0        & 85.9        & 55.5      & 42.8   & 17.4 & 33.6                     \\
                            & A.S.R.  & 78.2       & 59.0       & 99.8       & 98.8      & 71.6        & 51.4        & 64.7      & 85.1   & 94.1 & 58.9                    \\
\midrule                        
\multirow{2}{*}{Spectral}   & Acc.    & 98.1       & 97.6       & 68.3       & 64.1       & 85.9        & 83.4        & 40.9     & 44.0  & 62.2 & 20.5                      \\
                            & A.S.R.  & \textbf{0.1}       & \textbf{0.3}       & 3.1      & 56.4       & \textbf{0.2}        & \textbf{0.1}        & 88.4      & 83.2 & 5.2 & 83.5                       \\
\midrule 
\multirow{2}{*}{BaFFLe}     & Acc.    & 97.5       & 97.0      & 68.9       & 67.1       & 86.2        & 85.2        & 57.3     & 49.7 & 59.6 & 58.6                      \\
                            & A.S.R.  & 61.8       & 33.2       & 98.2       & 89.2       & 53.8       & 28.0        & 60.9      & 74.6 & 4.3 & 2.7                       \\
\midrule 
\multirow{2}{*}{IRLS}       & Acc.    & 97.8       & 97.2       & 68.9       & 64.1       & 86.8        & 85.8        & 80.8     & 78.9  & 13.7 & 13.0                     \\
                            & A.S.R.  & 47.2       & 55.5       & 98.1       & 98.6       & 35.7        & 51.3       & 8.4      & 6.1      & 95.8 & 97.4                 \\
\midrule 
\multirow{2}{*}{RLR}       & Acc.    & 97.6       & 96.3      & -      & -     & 85.9        & 84.4        & 55.2      & 43.4     & 56.3 & 50.2                  \\
                            & A.S.R.  & 4.2       & 1.0       & -      & -       & 0.8        & 4.6       & 65.0      & 84.7 & 18.6 & 27.2                       \\
\midrule 
\multirow{2}{*}{Ours}       & Acc.    & 98.1       & 97.5       & 68.2       & 63.1       & 86.9        & 85.9       & 79.9      & 74.3   & 61.3 & 55.7                     \\
                            & A.S.R.  & \textbf{0.2}       & 2.3      & \textbf{2.6}      & \textbf{1.0}       & \textbf{0.2}        & \textbf{0.2}        & \textbf{4.9}      & \textbf{1.6}         & \textbf{2.2} & \textbf{2.2}               \\
\bottomrule
\end{tabular}
\end{table*}
Table~\ref{table:detection} shows the comparison results of the detection performance between Spectral and our approach over the benchmark datasets for both IID and Non-IID data settings. 
Our approach outperforms the 
Spectral method 
across different benchmark tasks. 
\tdsc{Specifically, the proposed approach identifies malicious clients more accurately and stably 
with 0.014 FPR, 0 FNR, and 0.982 F1-score for IID data distribution and 0.016 FPR, 0.006 FNR, and 0.97 F1-score for Non-IID data distribution on average 
across all benchmark tasks.
Compared with Spectral, our approach achieves an average improvement of 35.4\% and 40.2\% on detection accuracy in terms of F1-score for IID and Non-IID data.
With regard to FPR and FNR, our approach realizes an average reduction of 21.8\%/23.4\% in FPR and 15\%/14.4\% reduction in FNR than the Spectral method for IID and Non-IID data, respectively. }


\tdsc{We observe that, 
the FNR of Spectral detection increases to 12\% for CIFAR-10 and 15\% for Yahoo Answers when dealing with Non-IID training data.
We hypothesize that the issue primarily stems from the model update representation problem in Spectral when dealing with complex model architectures, and the limitations in handling challenging scenarios where Non-IID data results in significant parameter updates.
Besides, since the Spectral adopts a dynamic average threshold, it will identify clients with reconstruction errors larger than the average one as anomalies in a ``hard'' binary manner,
which leads to the high rate of false-positive alarms, 23.2\%/25\% on average for IID/Non-IID.}

\tdsc{Figures~\ref{fig:baffle_iid} and \ref{fig:baffle_noniid} show the comparison results of detection accuracy in term of F1-score between BaFFLe and our approach. 
As stated before, 
the detection of BaFFLe is limited to identifying the global model as malicious or not, i.e., binary answer (1 or 0), instead of identifying the specific (and the number of) malicious clients.
We then calculate the F1-score across 15 settings for IID and Non-IID data.
Overall, the results show that our approach presents better detection accuracy on the benchmark tasks and achieves an average improvement of 15\% and 20\% in F1-score across the benchmark for IID and Non-IID data, respectively.}

In general, BaFFLe demonstrates unstable performance across the benchmark tasks. 
It presents better detection accuracy 
in the 
text classification tasks.
This is because BaFFLe relies on the prediction error variation for backdoor detection, making it fully constrained when  the prediction accuracy over validation data is well-preserved by the 
attackers~(e.g., the backdoor case of image classification tasks).
The adopted backdoor attack for the text classification task would cause a more obvious change to the prediction error on the validation data, which leads to 
better detection performance for AG-NEWS and Yahoo Answers datasets.
In addition, BaFFLe eliminates the backdoor injection
by rejecting global updates instead of aggregating over benign clients, which will lead to extra learning rounds and computing resource consumption. 
\begin{tcolorbox}[size=title]
{\textbf{Answer to RQ1: Our approach can effectively detect malicious clients for both IID and Non-IID data, demonstrating better detection accuracy than the baseline detection approaches.}}
\end{tcolorbox}




\subsection{RQ2: Can our approach effectively defend the global model against continuous backdoor attacks in runtime deployment?}
\label{subsec:defense}


To assess the effectiveness of our approach when deployed as a runtime defense mechanism, we conduct comparison experiments with the other defense methods, federated averaging~(Average), Spectral, BaFFLe, 
IRLS, and RLR in the attack scenario of naive approach~(i.e., attack in continuous rounds).


\tdsc{We evaluate the defense performance of our approach and the baseline methods under different attacker ratios varying from 10\% to 40\% (step length as 10\%).
For image datasets, we perform 20 training rounds and start the attack at round 10.
For text classification datasets, we start the attack at round 18 and 20 for AG-NEWS and Yahoo Answers tasks, respectively, as the backdoor embedding is much easier in this case.}
To accommodate the BaFFLe method, we enable the defense after the first 10 rounds to build a decent size of history global models as required by BaFFLe.
The evaluation metrics are attack success rate~(A.S.R.) concerning backdoor elimination
and prediction accuracy~(Acc.) on test data, so as to reflect the defense effectiveness against backdoor attacks as well as the prediction performance preservation on normal data.

\begin{figure*}[t]
    \centering
    \begin{subfigure}{0.19\textwidth}
    \centering
    \includegraphics[width=\textwidth]{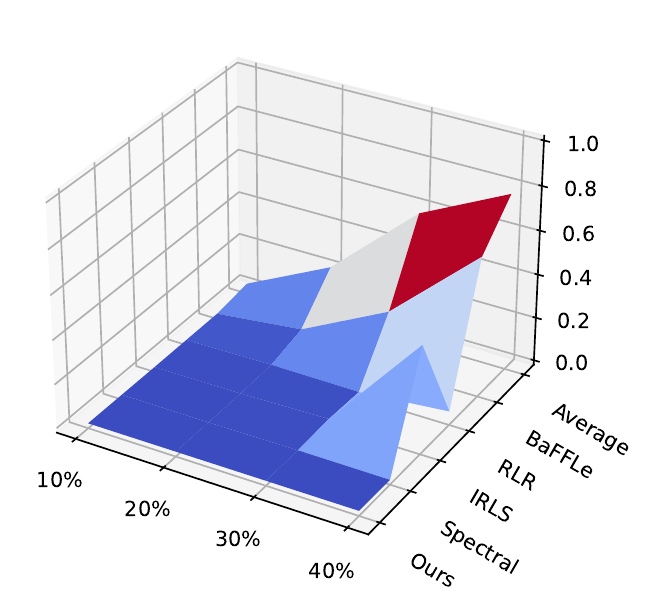} 
    \caption{MNIST}
    \label{fig:mnist_iid}
    \end{subfigure}
    \hfill
    \begin{subfigure}{0.19\textwidth}
    \centering
    \includegraphics[width=\textwidth]{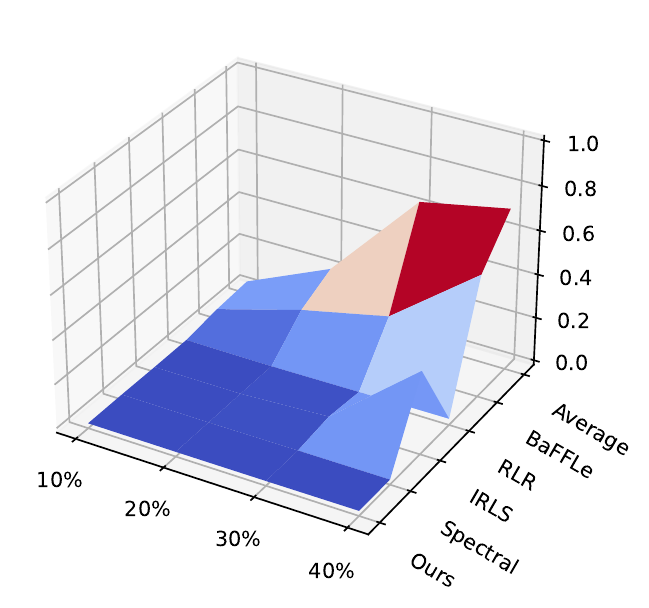}
    \caption{FASHION}
    \label{fig:fashion_iid}
    \end{subfigure}
    \hfill
    \begin{subfigure}{0.19\textwidth}
    \centering
    \includegraphics[width=\textwidth]{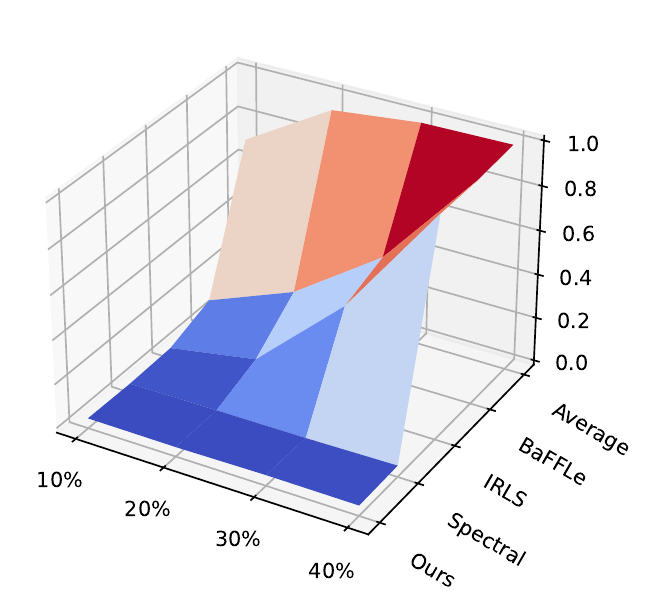} 
    \caption{CIFAR}
    \label{fig:cifar_iid}
    \end{subfigure}
    \hfill
    \begin{subfigure}{0.19\textwidth}
    \centering
    \includegraphics[width=\textwidth]{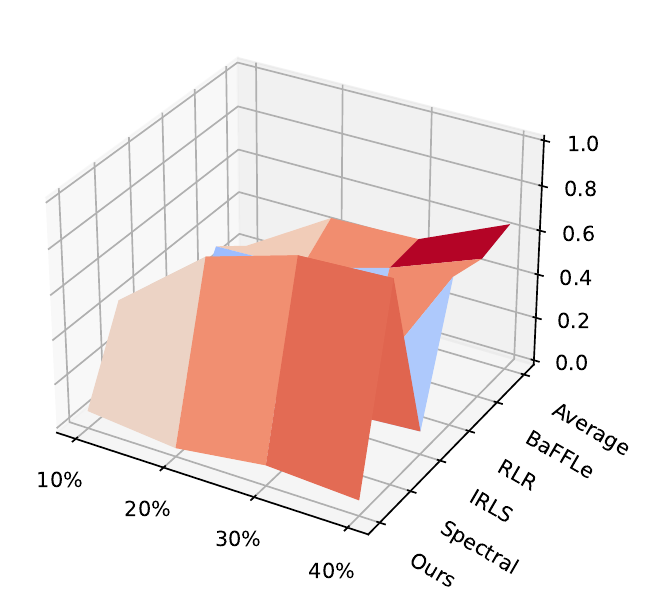} 
    \caption{AG}
    \label{fig:ag_iid}
    \end{subfigure}
    \hfill
    \begin{subfigure}{0.19\textwidth}
    \centering
    \includegraphics[width=\textwidth]{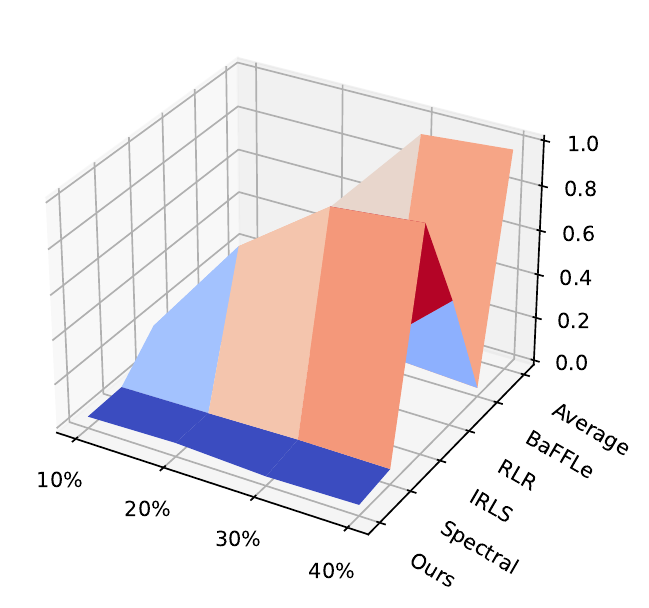} 
    \caption{YAHOO}
    \label{fig:yahoo_iid}
    \end{subfigure}
    \caption{\tdsc{Defense results in terms of attack success rate on the benchmark datasets (IID).}}
    \label{fig:defense_iid_all}
\end{figure*}

\begin{figure*}[t]
    \centering
    \begin{subfigure}{0.19\textwidth}
    \centering
    \includegraphics[width=\textwidth]{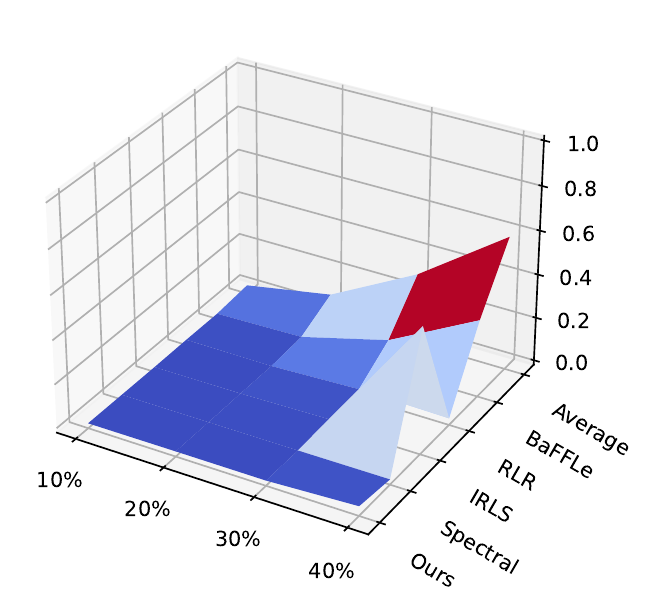} 
    \caption{MNIST}
    \label{fig:mnist_noniid}
    \end{subfigure}
    \hfill
    \begin{subfigure}{0.19\textwidth}
    \centering
    \includegraphics[width=\textwidth]{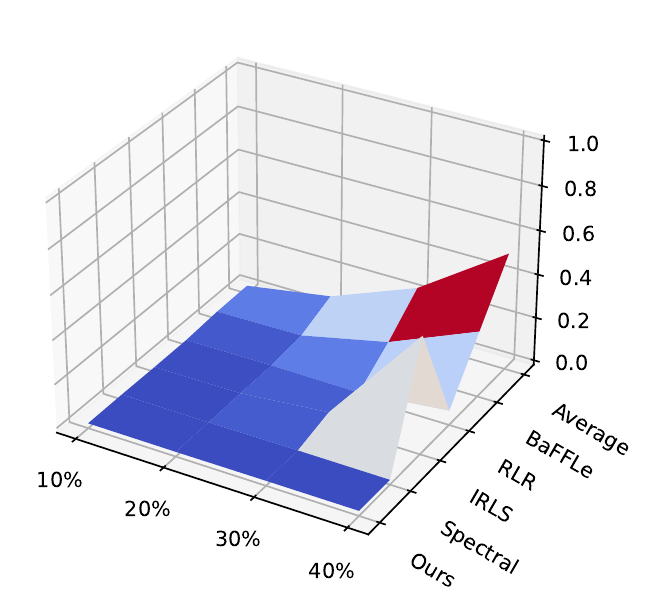}
    \caption{FASHION}
    \label{fig:fashion_noniid}
    \end{subfigure}
    \hfill
    \begin{subfigure}{0.19\textwidth}
    \centering
    \includegraphics[width=\textwidth]{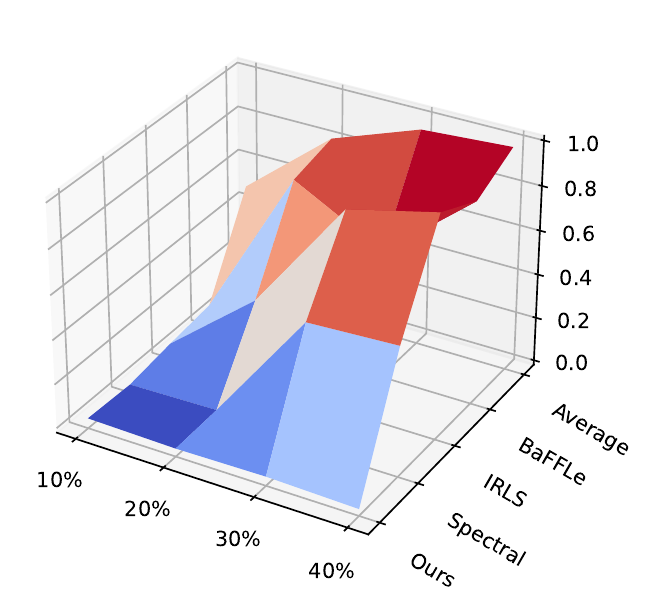} 
    \caption{CIFAR}
    \label{fig:cifar_noniid}
    \end{subfigure}
    \hfill
    \begin{subfigure}{0.19\textwidth}
    \centering
    \includegraphics[width=\textwidth]{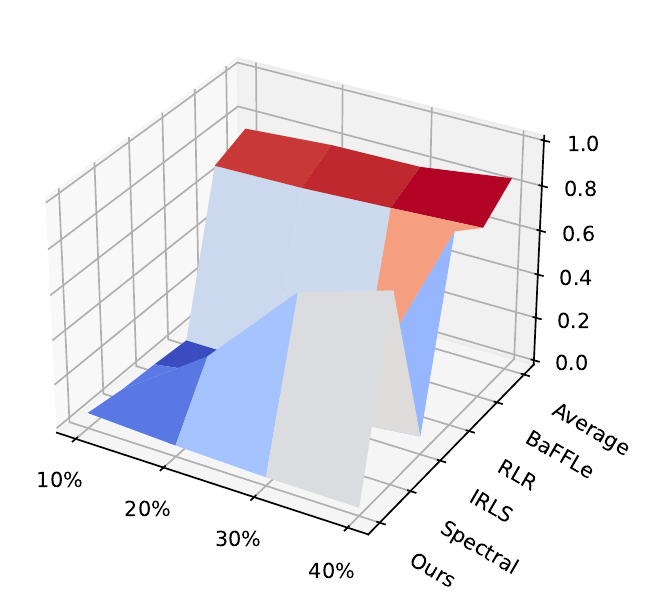} 
    \caption{AG}
    \label{fig:ag_noniid}
    \end{subfigure}
    \hfill
    \begin{subfigure}{0.19\textwidth}
    \centering
    \includegraphics[width=\textwidth]{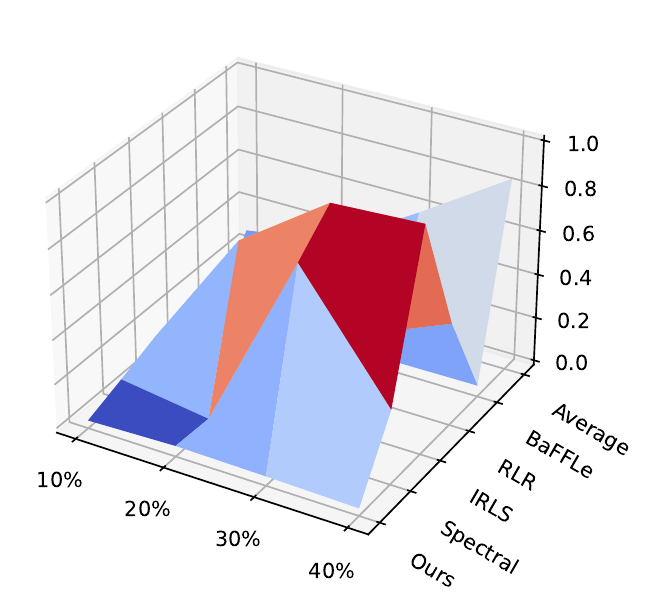} 
    \caption{YAHOO}
    \label{fig:yahoo_noniid}
    \end{subfigure}
    \caption{\tdsc{Defense results in terms of attack success rate on the benchmark datasets (Non-IID).}}
    \label{fig:defense_noniid_all}
\end{figure*}

\tdsc{The complete defense results of different approaches are presented in
Fig.~\ref{fig:defense_iid_all} and Fig.~\ref{fig:defense_noniid_all}.
The ideal result is to eliminate the backdoor injection which is reflected by low attack success rate.
We can see that our approach achieves effective and stable defense performance across the benchmark tasks for both IID and Non-IID data, outperforming the baseline methods or achieving comparable state-of-the-art results.}


\tdsc{Table~\ref{table:defense} summarizes the concrete comparison results in terms of prediction accuracy and attack success rate of different defense mechanisms across four benchmark tasks
when the attacker ratio is 40\%.
Overall, our approach achieves less than 2.6\% attack success rate on image classification tasks and less than 4.9\% on text classification tasks.
More specifically, compared with the federated averaging algorithm (\textit{Average}), our approach realizes an average reduction of 79.7\% and 69.2\% in attack success rate for IID and Non-IID data
across the benchmark datasets.}

\tdsc{Compared with the state-of-the-art robust aggregator IRLS,
our approach achieves a reduction of 55.0\% and 60.3\% on attack success rate for IID and Non-IID data.
Another state-of-the-art approach \textit{RLR} achieves defense by adjusting the update direction (i.e., signs) of model parameters, which achieves effectiveness for image classification tasks trained with convolutional network architectures.
However, it fails to converge when using ResNet models for image classification (CIFAR) and its effectiveness is compromised when dealing with natural language classification tasks where embedding layers are used in the model architectures, with 65.0\%/84.7\% A.S.R. for AG-NEWS and 18.6\%/27.2\% A.S.R. for Yahoo Answers in IID and Non-IID settings, respectively.}

\tdsc{BaFFLe reduces the attack success rate to a certain degree with an average reduction of 25.9\% and 25.1\% for IID and Non-IID data.
Still, the adversary attains more than 45\% attack success rate across the benchmark with BaFFLe as defense.
Spectral achieves less than 0.3\% attack success rate for MNIST and Fashion-MNIST.
However, its defense ability drops sharply~(56.4\% A.S.R. for CIFAR-10) when the model architecture becomes more complicated with heterogeneous data (Non-IID). 
In addition, Spectral fails to defend models processing sequential inputs where the attack success rate reaches an average of 85.8\% for AG-NEWS and 44.4\% for Yahoo Answers.
IRLS shows its effectiveness in defending the
text classification model for the AG-NEWS task, 
achieving 
56.3\% / 79.0\% (IID / Non-IID) reduction in A.S.R.
Still, IRLS fails to achieve effective defense for the LSTM model in the Yahoo Answers task, especially with a higher attacker ratio, 
laying up backdoor vulnerabilities.}

\tdsc{Moreover, our approach 
preserves 
the
prediction accuracy of the global models for all tasks.
The achieved test accuracy are 97.8\%, 65.7\%, 86.4\%, 77.1\% and 58.5\% for MNIST, CIFAR-10, Fashion-MNIST, AG-NEWS and Yahoo Answers datasets on average across different data distributions.
We can see that the achieved accuracy is 
about the same as 
the 
\textit{Average} algorithm with 97.4\%, 67.2\% and 87.0\% test accuracy for the image classification datasets.
As the backdoor used for text classification tasks along with a high attacker ratio causes more negative impacts on test accuracy, our approach 
achieves about 28.0\% and 33.0\% improvement in test accuracy for AG-NEWS and Yahoo Answers, respectively, than the \textit{Average} method.}




\begin{figure}[t]
    \centering
    \begin{subfigure}{0.47\columnwidth}
        \centering
        \includegraphics[width=\textwidth]{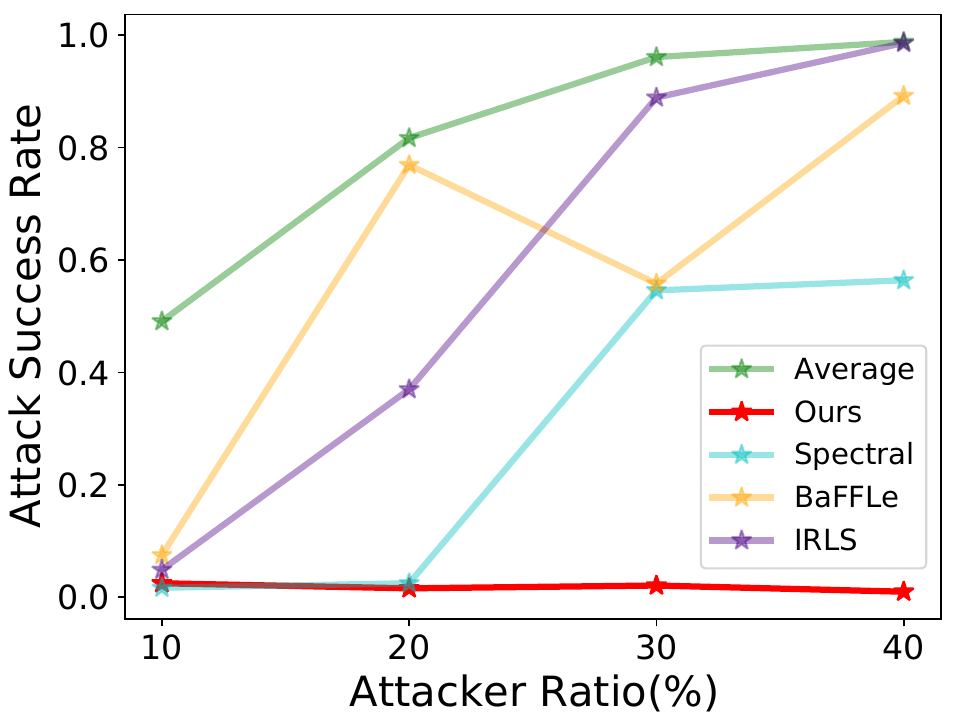}
        \caption{Across attacker ratios.}
        \label{fig:ASR}
    \end{subfigure}
    \begin{subfigure}{0.47\columnwidth}
        \centering
        \includegraphics[width=\textwidth]{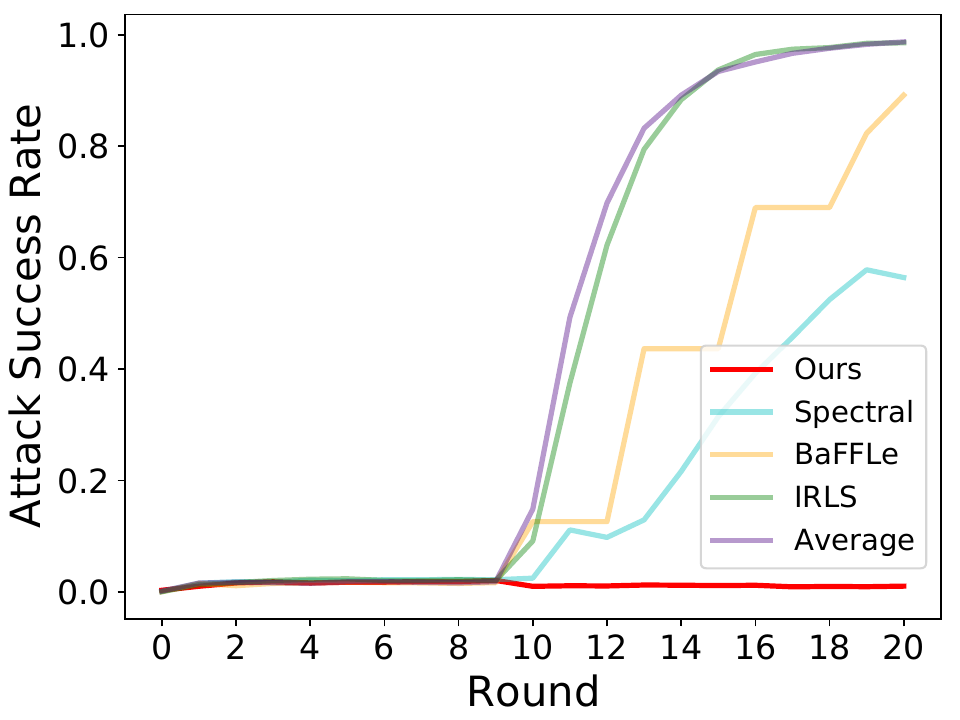}
        \caption{Across training rounds.}
        \label{fig:trace-ASR}
    \end{subfigure}
    \caption{Attack success rate attained by defense mechanisms across attacker ratios and training rounds.}
    \label{fig:asr}
\end{figure}

Fig.~\ref{fig:ASR} shows the attack success rate attained by different defense approaches under attacker ratios from 10\% to 40\% for CIFAR-10 in the Non-IID setting. 
The comparison results demonstrate the advantage of our approach in both defense effectiveness and stability across all attacker ratios.
Specifically,
our approach achieves stable and effective backdoor elimination, 
reducing the attack success rate to 1.8\% on average across all attacker ratios. 
Spectral achieves comparable attack success rate reduction when the attacker ratio is no more than 20\%. 
However, 
it loses the defense stability 
when the attacker ratio increases (54.6\%  A.S.R. for attacker ratio 30\% and 56.4\% A.S.R. for 40\%).
The other defense methods demonstrate effectiveness when the attacker ratio is 10\% but do not have sufficient stability in backdoor elimination as the attacker ratio increases.


Fig.~\ref{fig:trace-ASR} shows the traces of attack success rate across the training rounds
with different defense mechanisms 
under 40\% attacker ratio. 
We can see that our approach is the only one to succeed in defending the continuous backdoor attacks.
Compared with the Average algorithm, the other baseline methods are able to slow down the increase of attack success rate to some level, but still all reach high attack success rates higher than 50\%
when being attacked continuously.
\begin{tcolorbox}[size=title]
{\textbf{Answer to RQ2: Our approach can effectively and stably defend global models against continuous attacks, demonstrating 
better backdoor elimination than the baseline approaches.
}}
\end{tcolorbox}
\subsection{RQ3:
What is the performance of our approach under different backdoor patterns?}

\begin{table*}[t]
\caption{\tdsc{Defense results of accuracy (Acc.) and attack success rate (A.S.R.) of different defense mechanisms for logo backdoor patterns}\label{table:defense_patterns}}
\centering
\begin{tabular}{c|c|cc|cc|cc|cc}
\toprule
\multicolumn{2}{c|}{Tasks} &
\multicolumn{2}{c|}{MNIST (Copyright)}           & \multicolumn{2}{c|}{MNIST (Apple)}           & \multicolumn{2}{c|}{FASHION (Copyright)}   & \multicolumn{2}{c}{FASHION (Apple)}      \\ 
\cmidrule{1-10} 
Methods & Metric (\%)
& IID   & Non-IID & IID   & \multicolumn{1}{c|}{Non-IID} & IID   & Non-IID & IID   & Non-IID \\ \midrule
\multirow{2}{*}{Average}    & Acc.   & 97.8      & 97.4       & 98.0       & 97.3       & 86.8        & 86.4        & 86.8      & 86.7                        \\
                            & A.S.R.  & 60.2       & 17.0       & 83.8       & 36.3      & 52.9        & 38.3        & 79.7      & 50.7                     \\
\midrule                        
\multirow{2}{*}{Spectral}   & Acc.    & 97.8       & 97.5       & 98.1       & 97.6       & 86.6        & 83.8        & 86.7     & 85.1                       \\
                            & A.S.R.  & 0.2       & 0.2       & 0.2      & 0.1       & 0.2        & 0.1        & 0.2      & 0.1                        \\
\midrule 
\multirow{2}{*}{BaFFLe}     & Acc.    & 97.2       & 97.0      & 97.6       & 96.6       & 85.7        & 85.0        & 85.9     & 85.8                       \\
                            & A.S.R.  & 23.6       & 1.3       & 29.2       & 13.6       & 7.0       & 9.5        & 48.1      & 42.7                        \\
\midrule 
\multirow{2}{*}{IRLS}       & Acc.    & 97.8       & 97.8       & 97.9       & 97.4       & 86.7        & 86.2        & 87.1     & 86.6                      \\
                            & A.S.R.  & 4.3       & 1.2       & 12.6       & 34.6       & 30.3        & 28.3       & 65.4      & 46.7                    \\
\midrule 
\multirow{2}{*}{RLR}       & Acc.    & 95.8       & 96.7      & 97.0      & 96.3     & 85.7        & 85.1        & 86.4      & 85.3                       \\
                            & A.S.R.  & 4.4       & 1.5       & 4.5      & 3.7       & 3.1        & 1.4       & 2.5      & 6.7                      \\
\midrule 
\multirow{2}{*}{Ours}       & Acc.    & 98.1       & 97.7       & 97.9       & 97.5       & 86.8        & 84.1       & 86.4      & 81.7                        \\
                            & A.S.R.  & 0.1       & 0.1      & 0.2      & 0.1     & 0.2     & 0.1     & 0.1      & 0.2                    \\
\bottomrule
\end{tabular}
\end{table*}

\tdsc{To answer RQ3, we evaluate our approach against two practical backdoor patterns: \textit{Copyright} and \textit{Apple} logos from \cite{RLROzdayiKG21}.
These logos are \textit{gray-scale} images with size of 224$\times$244 and 200$\times$200, respectively, which are scaled appropriately for use as embedded backdoors.
As with the \textit{Square} pattern,
we investigate the effectiveness of our method against these practical backdoors, comparing it to other detection and defense techniques discussed in previous sections.
}


\tdsc{Table~\ref{table:defense_patterns} summarizes the defense results for the \textit{Copyright} and \textit{Apple} backdoor patterns.
The overall performance is consistent with what we observed for the previous backdoor pattern (Table~\ref{table:defense}). 
Our method achieves the lowest attack success rate across all tasks, with an average reduction of 52.3\% compared to the \textit{Average} algorithm, while preserving the prediction accuracy.
The \textit{Spectral} method shows comparable performance to ours.  
The other three baseline methods also reduce the attack success rate to some extent; however, for certain tasks (e.g., FASHION (Apple)), the attack success rate remains as high as 48.1\%, 65.4\%, and 6.7\%.
}

\begin{tcolorbox}[size=title]
{\tdsc{\textbf{Answer to RQ3: Our approach can effectively 
defend against malicious clients with different backdoor patterns like logo backdoor in practice.
}}}
\end{tcolorbox}

\subsection{RQ4:
What is the detection overhead when our approach is used as a runtime defense solution?}

\begin{table}[t]
\caption{\tdsc{Time consumption of defense mechanisms\label{table:time}}}
\centering
\begin{tabular}{c|c|c|c|c|c}
\toprule
Time(s) & MNIST   & CIFAR   & FASHION & AG\_NEWS & YAHOO\\ 
\midrule
Spectral & 0.06\tiny{+533}  & 0.08\tiny{+1098}  & 0.03\tiny{+527} & 0.04\tiny{+400}  & 0.05\tiny{+4528} \\ 
BaFFLe   & 17.64 & 57.48 & 17.91 & 86.68 & 354.20\\ 
IRLS    & 0.05  & 21.86 & 0.15  & 71.87  & 68.56\\ 
RLR     & 0.01  & -   & 0.01  & 1.36 & 1.22 \\
Ours     & 2.06  & 5.62 & 2.10  & 7.98  & 9.06\\
\bottomrule


\end{tabular}
\end{table}

Table~\ref{table:time} summarizes the time cost of different approaches when deployed as runtime defense.
For each approach, we calculate the average overhead with regard to the detection rounds, attacker ratios, and different data distribution types.
Among them all, Spectral has the best or comparable performance in runtime efficiency for all tasks.
This is because the anomaly detection model of Spectral is trained in advance based on the model updates obtained from each epoch in centralized training on test data.
We show the offline training time of the detection model of Spectral in the subscript.
RLR demonstrates competitive performance with Spectral on the benchmark tasks.
However, it achieves defense by gathering the client votes for model parameter sign change, therefore the defense overhead grows at least linear to the parameter size of the model architecture. 
As demonstrated in Section~\ref{subsec:defense}, RLR is also faced with the convergence problem when using complex models like ResNet architecture.

IRLS shows advantage in time cost for MNIST and Fahshion-MNIST tasks.
However, the IRLS algorithm needs to calculate the residuals of all parameters with regard to the linear regression line for all local models, which leads to cubical growth in time cost with the model parameter complexity.
The algorithm design of BaFFLe investigates the current global model's prediction error variation on the validation data against the history global models on each local client, which leads to the increase of time cost for backdoor detection.
\tdsc{Meanwhile, our approach achieves detection by inspecting the extracted RDM instead of the model parameters, which empowers our approach with the scalability to large model architectures.
For example, compared with IRLS, our approach achieves about 74.3\%, 88.9\% and 86.8\% improvement in runtime efficiency for CIFAR-10, AG-NEWS and Yahoo Answers tasks.
Our approach outperforms BaFFLe in time cost with 95.0\% improvement on average across all benchmark tasks.
Moreover, we could further reduce the detection overhead of our approach by preprocessing the data stimuli once and for all or parallelizing the process of model representation extraction.}

\begin{tcolorbox}[size=title]
{\textbf{Answer to RQ4: Our approach can be used as a runtime defense solution with sufficient efficiency, along with the scalability to complex model architectures.}}
\end{tcolorbox}

\xy{\subsection{Discussion}}
\tdsc{In this section, we evaluate the performance of our approach as the number of clients increases.
We further examine its effectiveness against varied attack configurations beyond adversary-selected backdoor injection and analyze the effect of sample size on detection accuracy. Lastly, we discuss the parameter settings of our approach and highlight key differences compared to previous methods.}

\begin{table*}[t]
\caption{\tdsc{Defense results of accuracy (Acc.) and attack success rate (A.S.R.) of different methods for 20 clients (IRLS and RLR result in Out of Memory errors)}\label{table:defense_client_size}}
\centering
\begin{tabular}{c|c|cccc|cccc}
\toprule
{Methods} & {Metric (\%)} &
\multicolumn{4}{c|}{IID} & \multicolumn{4}{c}{Non-IID} 
\\ 
\cmidrule{1-10} 
\multicolumn{2}{c|}{Attacker Ratio}
& 10\%   & 20\% & 30\%   & 40\% & 10\%   & 20\% & 30\%   & 40\% \\ \midrule
\multirow{3}{*}{Average}    & Acc.   & 44.4      & 22.7       & 13.5       & 10.5       & 47.4        & 29.6        & 16.8      & 10.5                       \\
                            & A.S.R.  & 27.1       & 84.7       & 95.5      & 99.7      & 21.3        & 68.4        & 90.2      & 99.7        
\\
                            & Time(s)  & 0.001       & 0.002       & 0.001       & 0.001      & 0.004        & 0.003        & 0.003      & 0.004     
                    \\
\midrule                        
\multirow{3}{*}{Spectral}   & Acc.    & 53.1       & 52.3       & 52.5       & 52.5       &  55.0       &    55.2     & 53.4    &  54.5                       \\
                            & A.S.R.  & 3.0       & 4.1       & 3.0      & 7.1       &     5.6    & 3.6        &   2.5    & 2.2  \\
                            & Time(s)\scriptsize{+4528}  & 0.051       & 0.079       & 0.054       & 0.048      &  0.052       & 0.073         & 0.066      & 0.089                          \\
\midrule 
\multirow{3}{*}{BaFFLe}     & Acc.    & 48.4       & 47.6      & 49.1       & 48.4       &   49.9      & 51.5        & 52.6     &   51.9                    \\
                            & A.S.R.  & 2.7       & 4.7      & 4.8       & 9.7       &    2.9    & 8.0        & 2.6     &    2.4
                             \\
                            & Time(s)  & 568.350       & 568.828       & 570.530       & 596.911      &   469.898      &  450.229        &  419.010     &   417.066  \\
\midrule 
\multirow{3}{*}{Ours}       & Acc.    & 53.0       & 53.0       & 53.1       & 54.1       &    54.3     & 56.1       & 55.9      &   55.2                    \\
                            & A.S.R.  & 3.3       & 3.7      & 3.5      & 4.2      &     1.8    &   3.6      & 3.4      &     3.8
                            \\
                            & Time(s)  & 20.415       & 26.230       & 26.614       & 19.658      &   21.608      &   27.786       &   30.367    & 19.875  \\
\bottomrule
\end{tabular}
\end{table*}

\tdsc{\subsubsection{Evaluation on Increased Clients}
To evaluate the performance of our approach on more clients, we conducted additional experiments in a 20-client setting.
Table \ref{table:defense_client_size} presents the results for the Yahoo Answers task under different attacker ratios.
Compared with the \textit{Average} algorithm, our method achieves an average reduction of 69.9\% in attack success rate and a 29.9\%  improvement in accuracy.}

\tdsc{
Compared with \textit{Spectral} and \textit{BaFFLe} methods, our approach demonstrates more consistent defense performance across all settings.
Specifically, under a 40\% attacker ratio in the IID setting, \textit{Spectral} shows an attack success rate of up to 7.1\%, while \textit{BaFFLe} permits up to 9.7\%. Moreover, \textit{Spectral} suffers from a high false positive rate and \textit{BaFFLe} relies on model rejection at the global level instead of at the individual client level; they both lead to lower test accuracy.}

\tdsc{In terms of time overhead, \textit{Spectral} requires 4528 seconds for anomaly detection model training and has a runtime overhead of 0.064 seconds, demonstrating high efficiency for runtime deployment.
In comparison with \textit{BaFFLe}, our method achieves over 20 times improvement in runtime efficiency.
We also run the \textit{IRLS} and \textit{RLR} algorithms for defense in the 20-client setting. However, they both encountered out-of-memory errors due to the need to handle 20 copies of the model weights.
}

\tdsc{Compraed with the 10-client setting, we observe that as the dataset is distributed across more clients, the global model converges more slowly, leading to lower accuracy at the same training round.
In terms of defense performance, our approach consistently maintains its effectiveness across different attacker ratios.
Regarding  time cost,
the runtime scales linearly with the number of clients.
This overhead can be reduced by parallelizing the RDM extraction process across all local clients.}
\subsubsection{Evaluation on Varied Attacks.}
\xy{
We evaluate our detection approach against three more attack configurations: (1) label-flipping attacks \cite{LabelFlipAttack} where attackers flip the labels of local training data from the source class to a target class; (2) semantic backdoor attacks \cite{bagda20a} by assigning an attacker-chosen label to data with certain features instead of modified training data with an attacker-chosen backdoor, e.g., cars painted in green labeled as birds;
(3) stealthy attacks \cite{FLAdvlens} where the adversary maintains the attack stealth by ensuring the malicious update to be close to the benign agents' updates. 
}

\begin{table}[t]
\caption{Detection performance against varied attack configurations \label{table:varied_attack}}
\centering
\begin{tabular}{c|c|ccc}
\toprule
Data Type               & Attack                          & FPR     & FNR    & F1    \\ 
\midrule
\multicolumn{1}{c|}{\multirow{3}{*}{IID}}     & Label-flip                   & 0    & 0      & 1    \\
\multicolumn{1}{c|}{}                         & Semantic               & 0    & 0      & 1    \\
\multicolumn{1}{c|}{}                         & Stealthy                   & 0    & 0      & 1     \\ 
\midrule
\multicolumn{1}{c|}{\multirow{3}{*}{Non-IID}} & Label-flip                  & 0.025    & 0      & 0.96       \\
\multicolumn{1}{c|}{}                         & Semantic                 & 0    & 0      & 1   \\
\multicolumn{1}{c|}{}                         & Stealthy                    & 0    & 0      & 1       \\ 
\bottomrule
\end{tabular}
\end{table}
\xy{
In label-flipping attacks, we flip the label of 1 to 7 for the MNIST dataset.
We perform the attack at round 20 with attacker ratio varied from 0\% to 40\% for both IID and Non-IID data.
We measure the detection performance in terms of three metrics: FPR, FNR, and F1 score.
Table~\ref{table:varied_attack} shows the average results of 5 experiments settings w.r.t. attacker ratio (0\%-40\%) for IID and Non-IID data distribution.
Our approach achieves accurate detection for IID data, and achieves 0.025 FPR and 0.96 F1-score for Non-IID data, demonstrating effective performance against label-flipping attacks.
In general, 
unlike backdoor attack, which is attained by mixed training objectives - prediction accuracy on normal data and target prediction for data with selected trigger, local training under label-flipping attacks
is based on a selected target class assigned to normal data of a certain source class.
Compared with backdoor injection,
such label flipping attack 
will lead to more obvious behavior differences 
on normal data from benign clients.
Still, Non-IID data poses challenges to the detection performance of our method, as heterogeneous data will cause more behavior difference between benign clients.}

\xy{Semantic backdoor attacks aim to cause model mis-classification on input data with certain features.
In contrast with backdoor injection, semantic backdoors can cause the model to misclassify inputs without additional modification. 
For the CIFAR-10 task, we follow \cite{bagda20a} and use three features as semantic backdoors: green cars (30 images in CIFAR-10), cars with racing stripes (21 images), cars with vertically striped walls in the background (12 images), and change their labels to bird. 
We evaluate the detection performance under different attacker ratios and report the average results in terms of three metrics.
As shown in Table \ref{table:varied_attack},
our approach achieves accurate detection for both IID and Non-IID data distribution.
Similar to label-flipping attacks, the selected semantic backdoors are part of features of normal training data on which malicious clients are trained with different target labels from benign clients, which leads to model behavior deviation.
}

\xy{To achieve stealthy poisoning attacks, we follow \cite{FLAdvlens} and add loss terms corresponding to accuracy on normal data and weight update statistics.
For stealthy poisoning on the Fashion-MNIST task, we select 10 data instances with desired target class (different from the ground-truth) as the adversarial objective. 
To ensure the malicious weight update to be close to that of benign clients, we constrain the malicious updates by adding the $\ell_{2}$ norm between malicious updates and the average benign updates of the previous iteration, i.e., $\Vert \delta^{t}_{mal}-
\overline{\delta}^{t-1}_{ben} \Vert$ into the malicious objective function.
The evaluation results against stealthy attacks are summarized in Table \ref{table:varied_attack}.
Our approach could accurately detect the stealthy poisoning attacks for both IID and Non-IID data distribution.
Although the stealthy poisoning attack takes the weight update distance from benign updates into consideration and incorporates $\ell_2$ distances in the loss function, our detection approach
does not rely on inspecting model updates. 
In fact, despite the constraints w.r.t. weight statistics, this additional term in the objective function leads to more behavior deviation from benign ones, which can be effectively detected by our approach.
}

\subsubsection{Effect of Sample Size}
\begin{figure}[t]
\centering
\includegraphics[width=0.7\columnwidth]{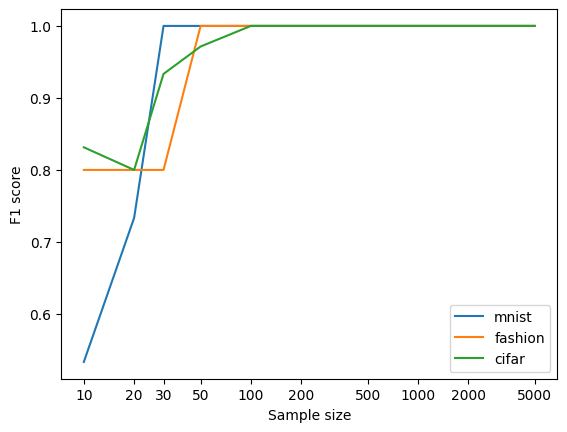} 
\caption{Detection accuracy under different sample size.}
\label{fig:sample_size}
\end{figure}

\xy{We demonstrate the effect of sample size on detection accuracy in the following.}
\xy{
Intuitively, a large sample size $n=m\cdot b
$ will capture more informative model representation, leading to more accurate detection, but at the cost of a higher computation overhead. 
To investigate the effect of sample size $n$ on detection accuracy, we perform experiments on image classification tasks and evaluate the detection accuracy by F1 score under different settings of $n$.
As shown in Figure~\ref{fig:sample_size}, the detection accuracy of our approach increases as the sample size increases until it reaches a plateau.
In our experiments, we follow a typical setting of 100 for each class
as a trade-off between effectiveness and efficiency.
}

\subsubsection{Hyperparameter Setting}
\tdsc{We set the LOF threshold of our approach to 1.5, which is generic and effective across all the benchmark tasks.
To accommodate the challenges brought by Non-IID data distribution with a high attacker ratio, we integrate a threshold refinement mechanism, as shown in Algorithm \ref{algo2}.
The threshold refinement mechanism relies on a dynamic thresholding strategy without a pre-fixed hyper-parameter, where the threshold $\delta_d$ is set as the mean value of the remaining clients' neighboring distance. 
Clients with higher neighboring distance than this threshold $\delta_d$ are deemed as malicious candidates. The LOF threshold $\delta_{re}$ is then refined to be the mean value of the LOF score of these potential malicious candidates. 
$\epsilon_d$ is used to avoid false positives in the case where all the remaining clients are benign ones. It is empirically set as the maximum neighboring distance among honest clients obtained in the early training rounds or the central aggregation server using test data. 
}


\tdsc{
Finally, we present a summary of the differences between the proposed approach and existing works in detection rationale and methodology.
The Spectral and BaFFLe techniques rely on inspecting model parameters and analyzing prediction error variation on validation data, respectively.
While the RLR method, relying on adjusting parameter signs, shows competitive performance in image classification tasks, it faces convergence issues, such as with the ResNet-18 model on CIFAR-10.
Our approach has fundamental differences in detection rationale and solution design:
(1) Detection rationale: Unlike existing methods that inspect local model parameters or assess prediction error variation, our approach uses sampling-based model representation.
This avoids the need for computations on large-size parameters in complex models and is effective against stealthy attackers that maintain high prediction accuracy.
(2) Solution design: We handle data heterogeneity and varying attacker ratios by using density-based metrics and proposing an iterative algorithm with a threshold refinement mechanism, rather than relying on mean or median of parameter updates. This makes our approach deliver consistent effectiveness against varied attack configurations.}


\section{Conclusion}
\label{sec:conclusion}
In this work, we propose a novel approach of runtime backdoor detection for federated learning~(FL).
We characterize the client behaviors by exploiting the model representation method in the form of
RDM, which provides a geometric representation of client models trained with different input statistics.
Then we perform the client differential analysis and quantify the client dissimilarity \textit{w.r.t.} their corresponding RDMs to capture the model deviation caused by backdoor injection.
We further propose an iterative algorithm to identify malicious clients and eliminate backdoor injection through outlier analysis based on the client dissimilarity measurements.
Our experimental evaluation demonstrates the effectiveness, stability and efficiency of our approach in defending FL against backdoor attacks across a range of tasks.



\ifCLASSOPTIONcompsoc
  \section*{Acknowledgments}
\else
  \section*{Acknowledgment}
\fi

This research was sponsored by the National Natural Science Foundation of China under Grant No. 62172019, 61772038,
and CCF-Huawei Formal Verification Innovation Research Plan; the Ministry of Education, Singapore under its Academic Research Fund under Grants Tier 1 21-SIS-SMU-033, Tier 3 MOET32020-0004; the National Research Foundation, Prime Ministers Office, Singapore under its National Cybersecurity R\&D Program (Award No. NRF2018NCR-NCR005-0001), NRF Investigatorship NRFI06-2020-0022-0001, the AI Singapore Programme (AISG2-RP-2020-019).
XZ received partial support from ELSA: European Lighthouse on Secure and Safe AI project (Grant No. 101070617 under UK guarantee).

\ifCLASSOPTIONcaptionsoff
  \newpage
\fi




\bibliographystyle{IEEEtran}
\bibliography{IEEEabrv,the.bib}
\newpage
\appendix

\begin{figure}[htb]
    \centering
    \begin{subfigure}{0.23\textwidth}
    \centering
    \includegraphics[width=\textwidth]{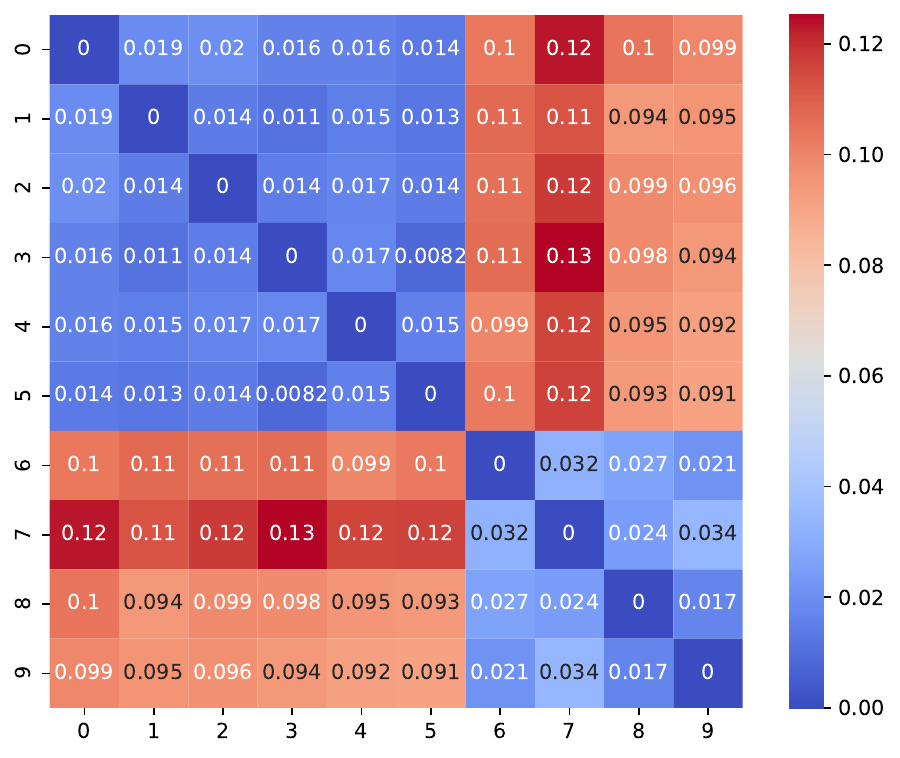} 
    \caption{MNIST (IID)}
    \label{fig:mnist_iid_heatmap}
    \end{subfigure}
    \hfill
    \begin{subfigure}{0.23\textwidth}
    \centering
    \includegraphics[width=\textwidth]{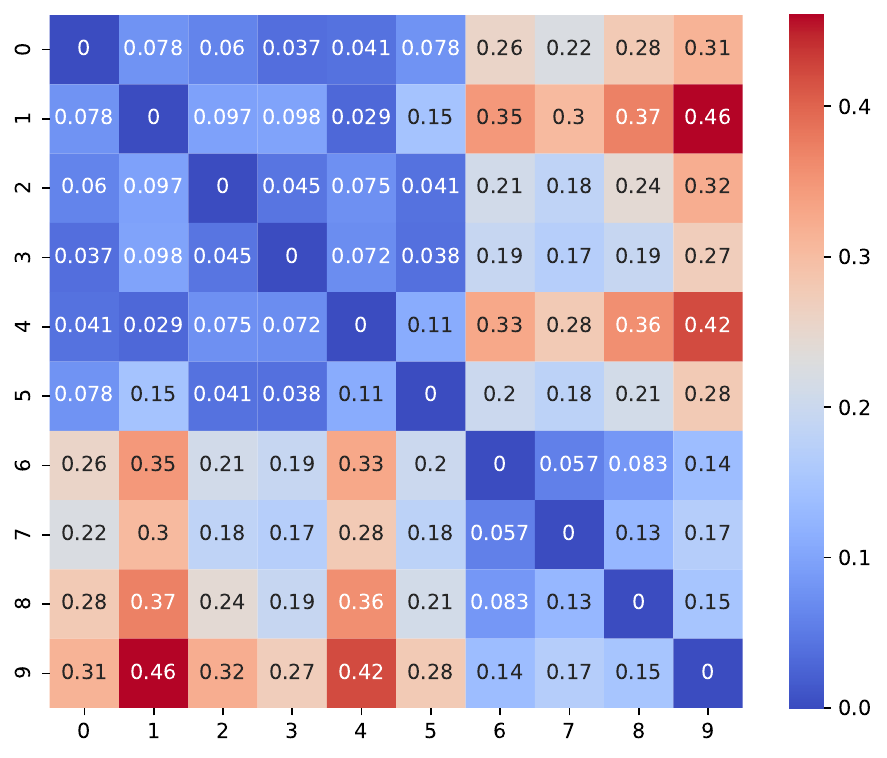}
    \caption{MNIST (Non-IID)}
    \label{fig:mnist_noniid_heatmap}
    \end{subfigure}
    \caption{\tdsc{Pairwise dissimilarity matrix for MNIST.}}
    \label{fig:dist_mat_mnist}
\end{figure}
\begin{figure}[htb]
    \centering
    \begin{subfigure}{0.23\textwidth}
    \centering
    \includegraphics[width=\textwidth]{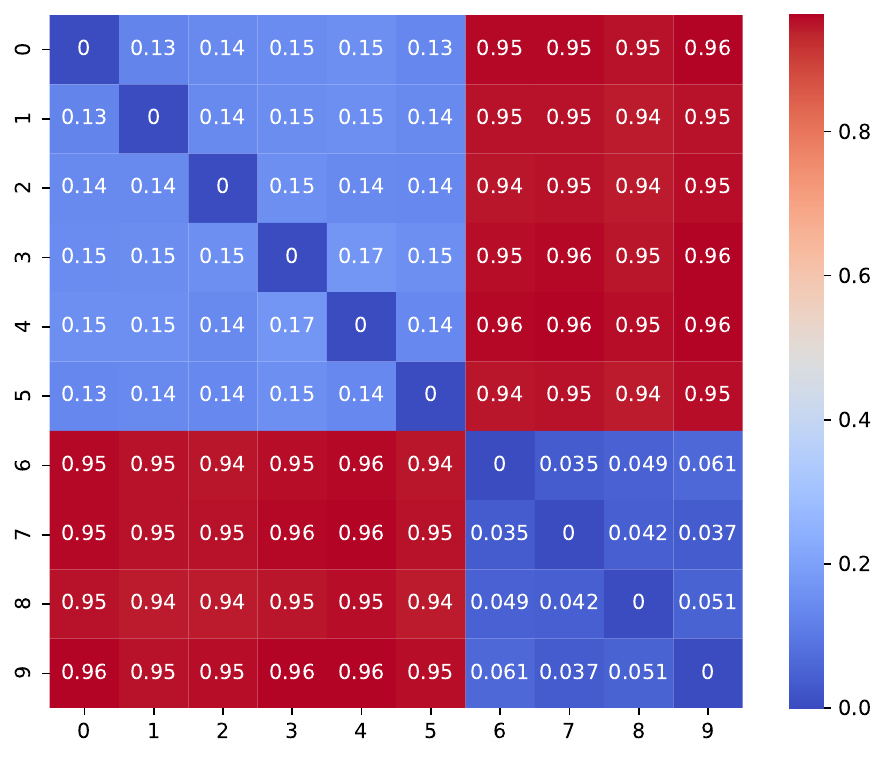} 
    \caption{YAHOO (IID)}
    \label{fig:yahoo_iid_heatmap}
    \end{subfigure}
    \hfill
    \begin{subfigure}{0.23\textwidth}
    \centering
    \includegraphics[width=\textwidth]{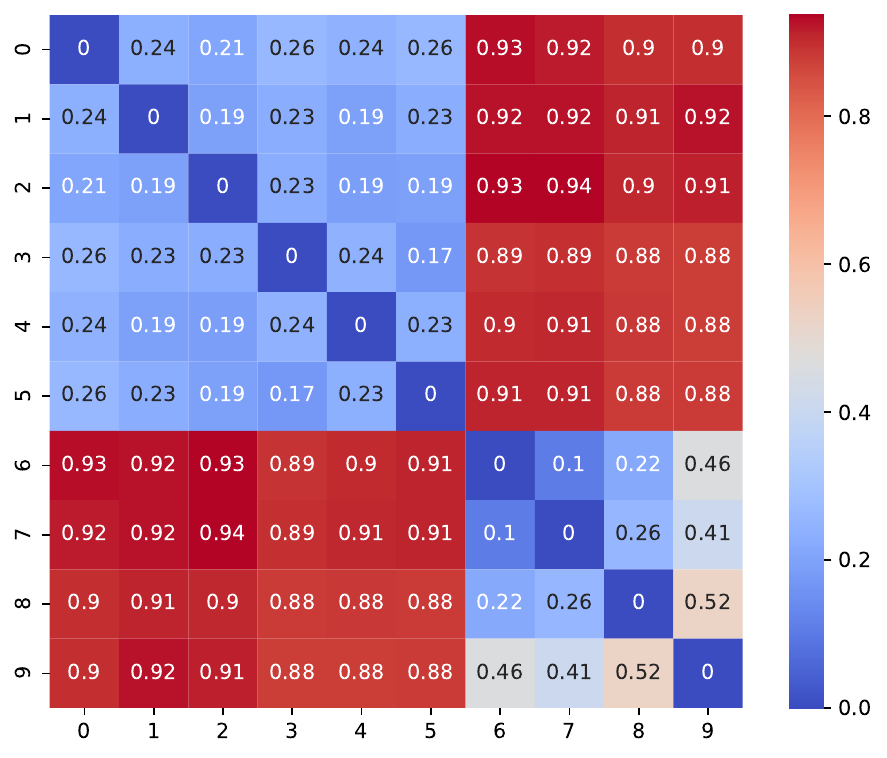} 
    \caption{YAHOO (Non-IID)}
    \label{fig:yahoo_noniid_heatmap}
    \end{subfigure}
    \caption{\tdsc{Pairwise dissimilarity matrix for Yahoo Answers.}}
    \label{fig:dist_mat_yahoo}
\end{figure}
\begin{table*}[htb]
    \centering
    \scriptsize
    \caption{\tdsc{Detection results on the benchmark datasets under different attacker ratios, attack rounds, and data distributions (F1-score for attacker ratios from 10\% to 40\%, FPR for attacker ratio 0\%)}}\label{tab:concrete_det_res}
    \begin{tabular}{cccccccccccccc}
        \toprule
        \multirow{4}{*}{\textbf{Dataset}} & \multirow{4}{*}{\textbf{Attacker Ratio}} & \multicolumn{12}{c}{\textbf{Attack Round}} \\ 
        \cmidrule(lr){3-14}
        & & \multicolumn{4}{c}{\textbf{Round 10}} & \multicolumn{4}{c}{\textbf{Round 20}} & \multicolumn{4}{c}{\textbf{Round 30}} \\
        \cmidrule(lr){3-6} \cmidrule(lr){7-10} \cmidrule(lr){11-14}
        & & \multicolumn{2}{c}{\textbf{IID}} & \multicolumn{2}{c}{\textbf{Non-IID}} & \multicolumn{2}{c}{\textbf{IID}} & \multicolumn{2}{c}{\textbf{Non-IID}}& \multicolumn{2}{c}{\textbf{IID}} & \multicolumn{2}{c}{\textbf{Non-IID}} \\
        \cmidrule(lr){3-4} \cmidrule(lr){5-6} \cmidrule(lr){7-8}
        \cmidrule(lr){9-10}
        \cmidrule(lr){11-12}
        \cmidrule(lr){13-14}
        & & \textbf{Spectral} & \textbf{Ours} & \textbf{Spectral} & \textbf{Ours} & \textbf{Spectral} & \textbf{Ours} & \textbf{Spectral} & \textbf{Ours} & \textbf{Spectral} & \textbf{Ours} & \textbf{Spectral} & \textbf{Ours}\\
        \midrule
        \multirow{5}{*}{MNIST}
        & 0\% ($\downarrow$) & 0.3 & 0 & 0.4 & 0 & 0.4 & 0 & 0.4 & 0.1 & 0.4 & 0 & 0.3 & 0\\
        & 10\%($\uparrow$) & 1 & 1 & 1 & 1 & 1 & 1 & 0.67 & 1 & 1 & 0.67 & 1 & 1\\
        & 20\%($\uparrow$) & 1 & 1 & 1 & 1 & 1 & 1 & 1 & 1 & 1 & 1 & 1& 1\\
        & 30\%($\uparrow$) & 1 & 1 & 1 & 1 & 1 & 1 & 1 & 0.86 & 1 & 1 & 1& 1\\
        & 40\%($\uparrow$) & 1 & 1 & 1 & 1 & 1 & 1 & 1 & 1 & 1 & 1 & 1 & 1\\
        \midrule
        \multirow{5}{*}{CIFAR} 
        & 0\% ($\downarrow$)& 0.3 & 0 & 0.4 & 0 & 0.3 & 0 & 0.4 & 0 & 0.4 & 0 & 0.4 & 0\\
        & 10\%($\uparrow$) & 0.4 & 1 & 0.4 & 1 & 0.5 & 0.67 & 0.67 & 1 & 0.5 & 1 & 0.67 & 1\\
        & 20\%($\uparrow$) & 1 & 1 & 0.8 & 1 & 0.8 & 1 & 1 & 1 & 1 & 1 & 1 & 0.8\\
        & 30\%($\uparrow$) & 0.75 & 0.86 & 0.57 & 0.8 & 0.86 & 0.86 & 1 & 1 & 1 & 1 & 0.57 & 0.86\\
        & 40\%($\uparrow$) & 0.89 & 1 & 0.86 & 1 & 1 & 1 & 0.86 & 1 & 1 & 0.89 & 0.75 & 0.89\\

        \midrule
        \multirow{5}{*}{FASHION} 
        & 0\% ($\downarrow$)& 0.6 & 0 & 0.5 & 0 & 0.4 & 0 & 0.4 & 0 & 0.4 & 0 & 0.4 & 0\\
        & 10\%($\uparrow$) & 0.4 & 1 & 0.4 & 1 & 0.5 & 1& 0.5 & 1 & 0.4 & 1 & 0.67 & 1\\
        & 20\%($\uparrow$) & 0.67 & 1 & 1 & 0.8 & 0.8 & 0.8 & 1 & 1 & 0.8 & 1 & 0.67 & 1\\
        & 30\%($\uparrow$) & 1 & 1 & 0.86 & 1 & 1 & 1 & 1 & 1 & 1 & 1 & 0.86 & 1\\
        & 40\%($\uparrow$) & 1 & 1 & 1 & 1 & 1 & 1 & 1 & 0.89 & 1 & 1 & 0.89 & 0.89\\

        \midrule
        \multirow{5}{*}{AG-NEWS} 
        & 0\% ($\downarrow$)& 0.4 & 0 & 0.5 & 0 & 0.6 & 0 & 0.5 & 0 & 0.3 & 0 & 0.4 & 0\\
        & 10\%($\uparrow$) & 0 & 1 & 0.33 & 1 & 0 & 1 & 0.33 & 1 & 0.33 & 1 & 0.4 & 1\\
        & 20\%($\uparrow$) & 0 & 1 & 0.29 & 1 & 0 & 1 & 0.29 & 1 & 1 & 0.8 & 0.4 & 1\\
        & 30\%($\uparrow$) & 0 & 1 & 0.25 & 1 & 0 & 1 & 0.25 & 1 & 0 & 1 & 0.25 & 1\\
        & 40\%($\uparrow$) & 0 & 1 & 0.25 & 1 & 0 & 1 & 0.25 & 1 & 0.89 &1 & 0.25 & 1\\

        \midrule
        \multirow{5}{*}{YAHOO} 
        & 0\% ($\downarrow$)& 0.4 & 0 & 0.4 & 0 & 0.3 & 0& 0.4 & 0 & 0.4 & 0 & 0.4 & 0\\
        & 10\%($\uparrow$) & 0.67 & 1 & 0.4 & 1 & 0.67 & 1 & 0.4 & 1 & 0.67 & 1 & 0.5 & 1\\
        & 20\%($\uparrow$) & 1 & 1 & 0.33 & 1 & 1 & 1 & 0.57 & 1 & 0.8 & 1 & 0.57 & 1\\
        & 30\%($\uparrow$) & 1 & 1 & 0.75 & 1 & 1 & 1 & 0.57 & 1 & 1 & 1 & 0.75 & 1\\
        & 40\%($\uparrow$) & 1 & 1 & 0.44 & 1 & 1 & 1 & 0.67 & 1 & 1 & 1 & 0.67 & 1\\

        \bottomrule
    \end{tabular}
\end{table*}

\tdsc{\textbf{Client Dissimilarity Quantification.}
Given the high dimensionality of the extracted client RDMs, we provide the pairwise client dissimilarity heatmap to illustrate the distances between benign and malicious models in terms of how they encode and recognize the input data.
As examples, we present the pairwise dissimilarity heatmaps for MNIST and Yahoo Answers datasets under both IID and Non-IID settings in Figure \ref{fig:dist_mat_mnist} and \ref{fig:dist_mat_yahoo}.
In these examples, clients 0-5 are benign models, and clients 6-9 are malicious.
As shown in Figure \ref{fig:dist_mat_mnist} and \ref{fig:dist_mat_yahoo}, 
the distances between benign models tend to be depicted in blue, indicating a high level of similarity in their data recognition and suggesting they form a close neighborhood where they serve as each other's $k$-distance neighbors (top left region in blue). 
In contrast, the RDM distances between  benign and malicious models are represented in red, indicating greater dissimilarity (top right and bottom left regions).
While the RDMs of malicious clients also share similarities due to their common backdoor injection objective, their distances exhibit greater variance. 
This suggests that malicious models are more sparsely distributed and located at varying distances from the densely clustered benign models. 
}

\tdsc{\textbf{Effectiveness of Backdoor Detection.}
Table \ref{tab:concrete_det_res}
summarizes the detailed detection results of our approach and the baseline method \textit{Spectral} for the benchmark datasets under attacker ratios from 0\% to 40\% (step length of 10\%), attack rounds from 10 to 30 (step length as 10) for both IID and Non-IID data distributions. 
We report the F1 score for the attacker ratios from 10\% to 40\% (the higher, the better) and the FPR (False Positive Rate) for the attacker ratio of 0\% (the lower, the better).}

\end{document}